\documentclass[11pt,twoside]{article}

\usepackage[english]{babel}
\usepackage{csquotes}

\usepackage{amsmath,empheq}
\usepackage{listings}

\PassOptionsToPackage{obeyspaces}{url}
\usepackage{hyperref}
\usepackage{float}
\usepackage{flafter}
\usepackage{booktabs}  
\usepackage{tabularx}  
\usepackage[dvipsnames]{xcolor}
\usepackage{tikz}
\usetikzlibrary{positioning,fit,calc}
\tikzset{block/.style={draw,thick,text width=2cm,minimum height=1cm,align=center}, line/.style={-latex}}
\tikzset{block2/.style={draw,thick,text width=3cm,minimum height=1cm,align=center}, line/.style={-latex}}
\tikzset{block3/.style={draw,thick,text width=5cm,minimum height=1cm,align=center}, line/.style={-latex}}

\usepackage{tocloft}
\usepackage{setspace}

\usepackage{acro}
\DeclareAcronym{mas}{short = MAS , long  = monoalphabetic substitution}
\DeclareAcronym{colt}{short = COLT , long  = columnar transposition}
\DeclareAcronym{ct2}{short = CT2 , long  = CrypTool 2}
\DeclareAcronym{prng}{short = PRNG , long  = pseudo-random number generator}
\DeclareAcronym{sct}{short = SCT , long  = single-columnar transposition}
\DeclareAcronym{cpu}{short = CPU , long  = central processing unit}
\DeclareAcronym{gpu}{short = GPU , long  = graphics processing unit}
\DeclareAcronym{chc}{short = CHC , long  = CrypTool 2 hill climbing routine}

\usepackage{caption}
\DeclareCaptionFont{white}{\color{white}}
\DeclareCaptionFormat{listing}{\colorbox{gray}{\parbox{\textwidth}{#1#2#3}}}
\definecolor{bg}{RGB}{240,240,240}

\lstdefinestyle{customc}{
  belowcaptionskip=1\baselineskip,
  breaklines=true,
  frame=shadowbox,
  xleftmargin=\parindent,
  language=C,
  showstringspaces=false,
  basicstyle=\footnotesize\ttfamily,
  keywordstyle=\bfseries\color{green!40!black},
  commentstyle=\itshape\color{purple!40!black},
  identifierstyle=\color{blue},
  stringstyle=\color{orange},
  tabsize=3,
  postbreak=\mbox{\textcolor{red}{$\hookrightarrow$}\space},
  rulecolor=\color{black},
  backgroundcolor=\color{bg},  
}

\lstdefinestyle{customconsole}{
  belowcaptionskip=1\baselineskip,
  breaklines=true,
  frame=shadowbox,
  xleftmargin=\parindent,
  language=bash,
  showstringspaces=false,
  basicstyle=\footnotesize\ttfamily,
  keywordstyle=\bfseries\color{green!40!black},
  commentstyle=\itshape\color{purple!40!black},
  identifierstyle=\color{black},
  stringstyle=\color{orange},
  tabsize=3,
}

\usepackage{ulem}

\begin{document}
\hypertarget{Chapter_LightweightIntroductionCUDA}{}
\label{LIL_Chapter_LightweightIntroductionCUDA}
(\hyperlink{author_Miroslav-Dimitrov}{Miroslav Dimitrov, Bernhard Esslinger})

\begin{center}
\setstretch{1.1}\LARGE\textbf{CUDA Tutorial -- Cryptanalysis of Classical Ciphers Using Modern GPUs and CUDA}
\date{}
\end{center}



\lstset{escapechar=@,style=customc}


\section{Introduction}
\label{sec:introduction}

CUDA (formerly an abbreviation of Compute Unified Device Architecture) is a parallel computing platform and API model created by Nvidia allowing software developers to use a CUDA-enabled graphics processing unit (GPU) for general purpose processing.\footnote{CUDA performs well but CUDA-enabled GPUs are only available from Nvidia. A similar API is OpenCL (Open Computing Language) – an open standard from Khronos Group –  which allows writing programs that execute across heterogeneous platforms consisting of some \ac{cpu}, some \ac{gpu}, digital signal processors (DSPs), field-programmable gate arrays (FPGAs), and other processors or hardware accelerators.}

Throughout this tutorial, we introduce the CUDA concepts in an easy-to-grasp interactive way. However, to fully benefit from the tutorial, some basic prerequisites are desirable:

\begin{itemize}
\item{A basic familiarity with C, C++ or a similar language}
\item{A basic understanding of cryptology}
\item{A basic understanding of heuristic techniques}
\item{Possession of a CUDA capable device}
\end{itemize}

Section \ref{sec:practical_introduction} gives a brief overview of the graphical processing unit (GPU). In this section we  setup our programming environment and learn how to exploit the multi-core architecture of the GPU by solving some trivial problems.

Section \ref{sec:GPUSubstAnalysis} provides the concept of GPU threads and  analyzes how  modern GPUs could be beneficial in terms of performing automatic cryptanalysis on classical cryptosystems.  

Starting from scratch, we implement a complete stand-alone GPU tool for automatically decrypting ciphertexts (ciphertext-only attack) encrypted by \ac{mas}. Throughout this process, we will learn how to architect the tool, what optimizations could significantly empower our routines, why the choice of an adequate metaheuristic\footnote{\url{https://en.wikipedia.org/wiki/Metaheuristic}} is critical, and how to draw sketches to enlighten the design process, by proactively solving upcoming issues.

The thread-count limitation problem could be easily solved by organizing the threads in blocks, as shown in Section \ref{sec:CUDABlocks}. Then, we setup another, beside the metaheuristic choice, critical part of the cryptanalysis -- the \ac{prng} nested in the device itself. Further discussion why \ac{prng}s are significantly beneficial to the classical cryptanalysis could be found in subsections \ref{subsec:Implement} and \ref{sub:standalone}. Having this at hand, we can further optimize our cryptanalysis tool by using a stand-alone pool of pseudo-random numbers, which allows to develop a more flexible metaheuristics.

Section \ref{sec:CUDADebug} briefly discusses some basic CUDA debugging tools used to differentiate errors yielded by the host from those yielded by the device.  Furthermore, we sketch out a compact overview of the CUDA memory model and why using shared memory could significantly increase the speed of our future applications.

Then, Section \ref{sec:dynamic} briefly discusses the usage of dynamically allocated memory on the device, and more specifically, why it should be avoided. As an example, we show how this usage could be avoided by using predefined macros.

Section \ref{sec:principles_comparison} first summarizes the most critical improvements made throughout the tutorial. Then, by following the design principles outlined during the previous sections, a stand-alone CUDA application is constructed for automatic cryptanalysis of ciphertexts encrypted by \acl{sct} cipher. At the end, the provided CUDA application is compared with a state-of-the-art tool for cryptanalysis.

The final Section \ref{sec:summary} summarizes the critical paradigms that should be addressed in the process of designing a CUDA cryptanalysis tool. 

\textcolor{red}{Please note, that all CUDA examples, as well as all accompanying files needed for their compilation, are provided inside the \textbf{Repository} folder. It could be downloaded from \url{https://www.cryptool.org/assets/ctb/CUDA-Tutorial_Repository.zip}.}

\section{Practical introduction to GPU architecture}
\label{sec:practical_introduction}

This section gives an overview of the GPU architecture. The examples are accompanied by visual interpretations and illustrations. For the sake of comprehensibility, we purposely avoid some technical details.

What is the abstract model of a given GPU? Imagine you have a function $F$, which just prints \textbf{Hello!} on the screen. If you launch this function once, a single output row will be printed. This is not the same when using a GPU, which is a multi-core system and as such it possesses lots of parallel processors called CUDA cores. Each of these cores is similar to a computer processor. Taken this into account, one could be already wondering: What will happen if we launch the function $F$ on all CUDA cores supplied by a given GPU? How to program and launch the function $F$ on the GPU in the first place? Moreover, how to spread the work of a given problem to a multi-core architecture in an efficient and productive way? In this tutorial, we are going to answer all these questions. 

\begin{figure}[h]
\centering
\includegraphics[width=250px]{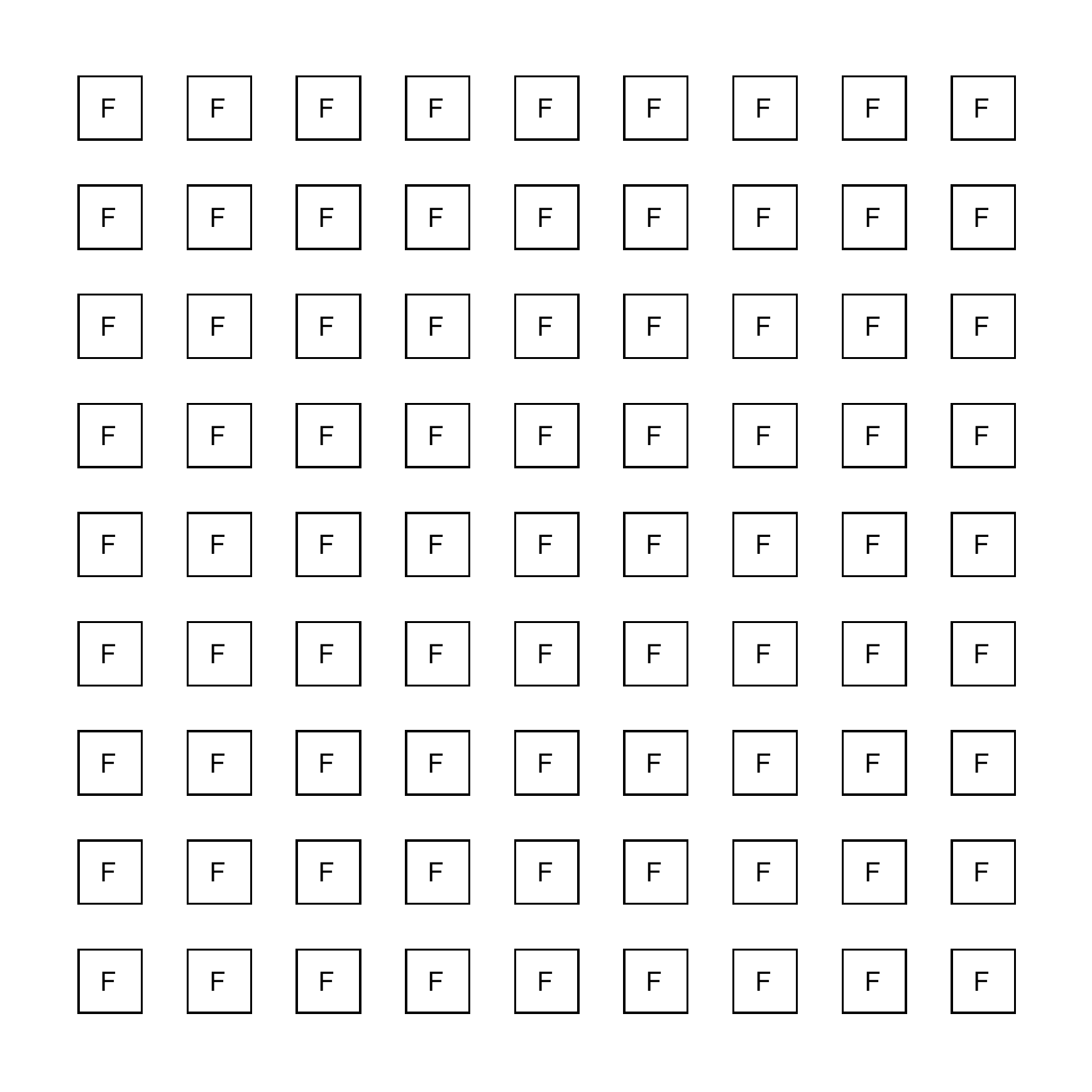} 
\caption{Example of GPU multi-core architecture}
\label{fig:ExampleCores} 
\end{figure}

Let's first construct a handy visual interpretation of a GPU multi-core architecture. Figure \ref{fig:ExampleCores} treats each square as a GPU core. In the current abstract example we have a total of 81 cores. Since we have initialized and loaded the function $F$ on each one of them, the GPU will launch 81 instances of the function $F$ and as a consequence, we will have a total of 81 printed rows with the message \textbf{Hello!}. 

Let's setup our environment and write down our very first CUDA program. For this, we first  need to download and install the CUDA Development Tools. For Windows, an installation guide can be found in \cite{cudawindows}, for Linux the corresponding installation guide can be found in \cite{cudalinux}. The Linux installation is pretty much straightforward. In contrast, for the Windows installation, we need to further install and link the C++ compiler tools. The following guide summarizes the major stages of setting up a Windows machine:

\begin{itemize}
\item{Install the CUDA driver.}

\item{Install the CUDA toolkit. Make sure the \textbf{nvcc} compiler is present by typing \textbf{nvcc --version} in the command prompt.}

\item{Make sure Visual Studio (\textbf{VS}) is present at the machine. Furthermore, make sure that the VS module \textbf{Desktop Development with C++} is installed. In case VS or the VS module is missing, you can always grab the free Visual Studio 2019 community version to be found in \cite{vstudio}.}

\item{Locate the \textbf{cl.exe file}. For example, if you are using Visual Studio 2019 community version, the default location of the file is similar to \path{C:\Program Files (x86)\Microsoft Visual Studio\2019\Community\VC\Tools\MSVC\<version>\bin\Hostx64\x64}.}

\item{Go to the directory where your CUDA examples are downloaded. In case you want to compile a file with name \textbf{example.cu}, the following command should be typed:



\begin{lstlisting}
nvcc -ccbin="C:\Program Files (x86)\Microsoft Visual Studio\2019\Community\VC\Tools\MSVC\<version>\bin\Hostx64\x64"  example.cu -o example
\end{lstlisting}

In case no compilation errors occur, the compiled file \textbf{example.exe} should be created in the current directory.} 

\end{itemize}

The CUDA platform is designed to work with programming languages such as, but not limited to, C, C++, and Fortran. Throughout this tutorial we are going to use the C language. 

Throughout the tutorial, by \textbf{host} we always reference to the CPU, while by \textbf{device} -- we reference to the GPU. Every CUDA program could be logically divided into two parts. The first part consists of the \textbf{host-related} code routines, i.e. the source code relevant to the host, while the second part contains the \textbf{device-related} code routines, i.e. the source code relevant to the external device. Assuming that our machine is prepared, let's compile our first CUDA application.

\begin{figure}[h]
\centering
\includegraphics[width=350px]{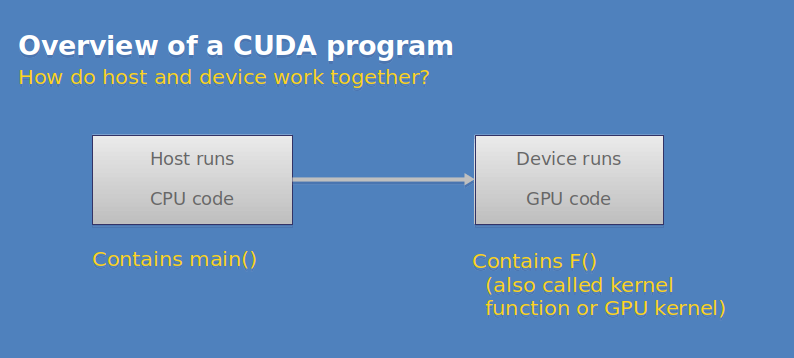} 
\caption{FlowChart: Overview of a CUDA program}
\label{fig:FlowChartOverview} 
\end{figure}

Figure \ref{fig:FlowChartOverview} shows how host and device codes works together.

\subsection{Our first CUDA programs}
\label{subsec:first_CUDA_programs}

\lstset{language=C}
\begin{minipage}{\linewidth}
\begin{lstlisting}[label={listing:1}, caption=Our first CUDA program, style=customc, numbers=left,stepnumber=1]
#include <stdio.h>

__global__ void F(){
	printf("Hello!\n");
}

int main() {
	F<<<1,81>>>();
	cudaDeviceSynchronize();
	return 0;
}
\end{lstlisting}
\end{minipage}

The function $F$ (see Listing \ref{listing:1}) is a function specific to the device only. By using the declaration \textbf{\_\_global\_\_} for some function $G$, we announce that $G$ is going to be executed on the device only. In CUDA, we have three different function declarations:

\begin{itemize}
\item{\textbf{\_\_global\_\_}: functions, that are called from the host and then executed on the device; they are called GPU \textbf{kernels}.}
\item{\textbf{\_\_device\_\_}: functions, that could be called only from the device and executed only on the device.}
\item{\textbf{\_\_host\_\_}: functions, that could be called only from the host and executed only on the host.}
\end{itemize}

However, as we will see later throughout the tutorial, for example in subsection \ref{sub:standalone}, in case we need a function to be shared with both the device and the host, we can use both \textbf{\_\_host\_\_} and \textbf{\_\_device\_\_} declarations.

The actual initialization of the program starts from the main() function. Then, we call the GPU kernel function $F$ by using this triple-bracket alike syntax: $<<<x,y>>>$, where $x$ defines the number of \textbf{blocks} (will be discussed in Section \ref{sec:CUDABlocks} ), and $y$ defines the number of threads (cores) per block. 

Since our hypothetical GPU example (see Figure \ref{fig:ExampleCores}) is based on exactly 81 cores, we launch the kernel by supplying a tuple $<<<1,81>>>$, which means that we are going to launch the $F$ function on all available 81 cores, without dividing them by blocks. Similarly to a C compiling procedure, the compilation is done by using \textbf{nvcc -o $\lbrace$programName$\rbrace$ $\lbrace$programName$\rbrace$.cu}. \textbf{NVCC} is the CUDA compiler driver.  Please note, that, as stated in the CUDA documentation, all non-CUDA compilation steps are forwarded to a C++ host compiler that is supported by nvcc.

The \textbf{cudaDeviceSynchronize()} function is a CUDA function, which handles the synchronization of all the threads we have started. Using synchronization is a good practice as it helps to properly print out the messages defined in the $F$ function. We want to emphasize that the kernel calls are asynchronous, which means that the control is returned to the host before the kernel completes. Having this in mind, the application could terminate before the kernel had the opportunity to print out the desired messages.

Let's compile and run the example. We can see that the message \textbf{Hello!} is printed 81 times in the console. However, in the given situation, we are not able to link a specific row with the corresponding parent GPU core. CUDA provides us with some very important built-in primitives, which can be utilized for thread's identification numbers. Let's slightly modify the example to illustrate this (see Listing \ref{list:threads}).

\begin{minipage}{\linewidth}
\begin{lstlisting}[label={list:threads}, caption=CUDA thread identification, style=customc, numbers=left,stepnumber=1]
#include <stdio.h>

__global__ void F(){
	printf("Hello from thread number %d!\n", threadIdx.x);
}

int main() {
	F<<<1,81>>>();
	cudaDeviceSynchronize();
	return 0;
}
\end{lstlisting}
\end{minipage}

The threadIdx.x number returns the unique ID of the thread. The IDs starts from 0, and by considering the current example, up to 80 (inclusive). Now, in case we want to find the squares of all the integer numbers from 0 to 80, we can change a little bit the function F to solve this problem in parallel, benefiting from the GPU architecture (see Listing \ref{list:squares}).

\begin{minipage}{\linewidth}
\begin{lstlisting}[label={list:squares}, caption=CUDA parallel squaring, style=customc, numbers=left,stepnumber=1]
#include <stdio.h>

__global__ void F(){
	printf("%d squared is: %d! \n", threadIdx.x, threadIdx.x*threadIdx.x);
}

int main() {
	F<<<1,81>>>();
	cudaDeviceSynchronize();
	return 0;
}
\end{lstlisting}
\end{minipage}

\subsection{Parallel squaring -- Part I}
\label{subsec:sqaring_filling_kernel_memory}

What if we want to square a set of numbers, which should be read from the host memory? What is the mechanism of transferring data from the host memory to the device memory? The procedure is similar to the memory allocation in a C language routine. We call the function \textbf{cudaMalloc((void **)\&P, SIZE)} which is going to reserve a memory block of size \textbf{SIZE} on  the device. The pointer to the reserved block will be saved to the pointer \textbf{P}.

In case we don't need this memory allocation anymore, we should free it with \textbf{cudaFree(P)}. To interchange memory blocks between the host and the device we use the built-in CUDA function \textbf{cudaMemcpy()}. Let's consider an example which illustrates the usage of these functions (see Listing \ref{list:ex1}). 

\begin{minipage}{\linewidth}
\begin{lstlisting}[label={list:ex1}, caption=CUDA parallel squaring by filling the kernel memory (one-way), style=customc, numbers=left,stepnumber=1]
#include <stdio.h>
#define threads 81

__global__ void H(int *d_seeds){
	int element = d_seeds[threadIdx.x];
	printf("%d squared is: %d \n", element, element*element);
}

int main() {

	int seeds[threads];
	
	for (int j=0; j<threads; ++j)
		seeds[j] = j;
		
	int size = sizeof(int);
	int *d_seeds;
	
	// allocate memory block on the device
	cudaMalloc((void **)&d_seeds, size*threads);	
	
	// copy the seeds array to the device memory
	cudaMemcpy(d_seeds,seeds,size*threads,cudaMemcpyHostToDevice);
	
	H<<<1,threads>>>(d_seeds);
	cudaDeviceSynchronize();
	cudaFree(d_seeds);
	return 0;
	
}
\end{lstlisting}
\end{minipage}

\begin{figure}[H]
\centering
\begin{tikzpicture}
  \node[block2] (ha) {empty};  
  \node[block,right=3cm of ha] (da) {empty};
  \node[draw,inner xsep=5mm,inner ysep=8mm,fit=(ha),label={60:CPU (host)}](f){};
  \node[draw,inner xsep=5mm,inner ysep=8mm,fit=(da),label={60:GPU (device)}]{};
\end{tikzpicture}
\caption{At the beginning, there is no data on both the CPU and the GPU. }
\label{fig:Ex1P1}
\end{figure}
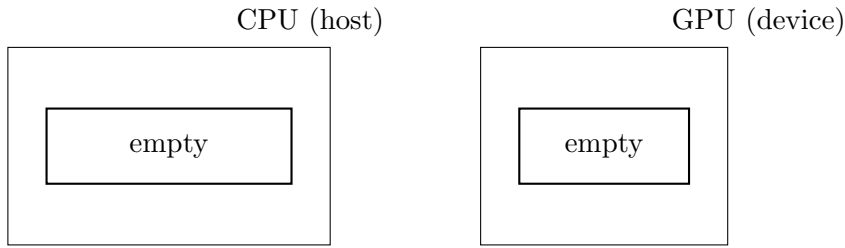

Now let us inspect the code line by line. Throughout the process, we are going to depict the memory blocks of the host and the device. As shown in Figure \ref{fig:Ex1P1}, at the beginning both the memory blocks, the host related one (on the left) and the device related one (on the right) are empty.

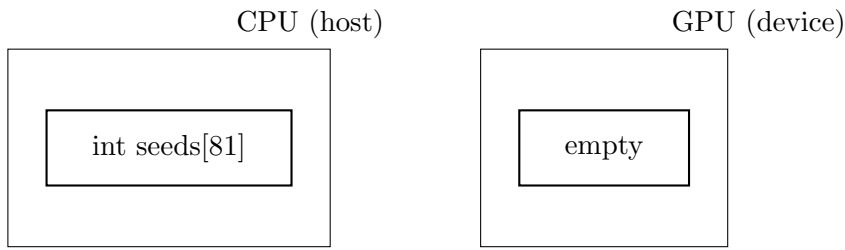
\begin{figure}[H]
\centering
\begin{tikzpicture}
  \node[block2] (ha) {int seeds[81]};
  \node[block,right=3cm of ha] (da) {empty};
  \node[draw,inner xsep=5mm,inner ysep=8mm,fit=(ha),label={60:CPU (host)}](f){};
  \node[draw,inner xsep=5mm,inner ysep=8mm,fit=(da),label={60:GPU (device)}]{};
\end{tikzpicture}
\caption{The integer array \textit{seeds} inside the host memory is initialized and filled with values.}
\label{fig:Ex1P2}
\end{figure}

At line 11 of Listing \ref{list:ex1} we declare an array of integers called \textbf{seeds}. We will use it as buffer located on the host memory. Then, at lines 13 and 14, we populate the cells in the  host memory block with the integer numbers from 0 to 80 (the total number of our threads). At this moment our memory snapshot of the host and the device is given in Figure \ref{fig:Ex1P2}.

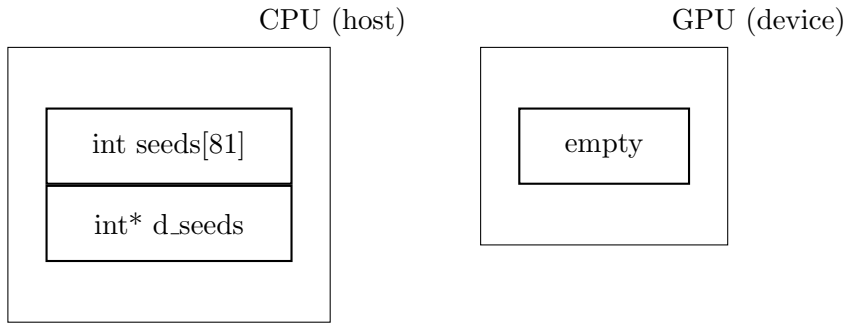
\begin{figure}[H]
\centering
\begin{tikzpicture}
  \node[block2] (ha) {int seeds[81]};
  \node[block2, below=0cm of ha] (hb) {int* d\_seeds};
  \node[block,right=3cm of ha] (da) {empty};
  \node[draw,inner xsep=5mm,inner ysep=8mm,fit=(ha) (hb),label={60:CPU (host)}](f){};
  \node[draw,inner xsep=5mm,inner ysep=8mm,fit=(da),label={60:GPU (device)}]{};
\end{tikzpicture}
\caption{The integer pointer \textit{d\_seeds} is created, which is going to point to a block in the device (GPU) memory.}
\label{fig:Ex1P3}
\end{figure}

At line 17 of Listing \ref{list:ex1} we  declare an integer pointer called \textbf{d\_seeds}, which is going to point to a memory block inside the device (not allocated yet). As shown in Figure \ref{fig:Ex1P3}, there is almost no change in the snapshot of the host and device memory blocks.

\begin{figure}[H]
\centering
\begin{tikzpicture}
  \node[block2] (ha) {int seeds[81]};
  \node[block2, below=0cm of ha] (hb) {int* d\_seeds};
  \node[block,right=3cm of ha] (da) {int[81]};
  \node[draw,inner xsep=5mm,inner ysep=8mm,fit=(ha) (hb),label={60:CPU (host)}](f){};
  \node[draw,inner xsep=5mm,inner ysep=8mm,fit=(da),label={60:GPU (device)}]{};
  \draw[red, ultra thick, -] (hb.east) -- (da.west);
\end{tikzpicture}
\caption{Using cudaMalloc() a memory block on the GPU is allocated to contain 81 integer cells. d\_seeds points to the exact memory address of this block inside the GPU memory.}
\label{fig:Ex1P4}
\end{figure}
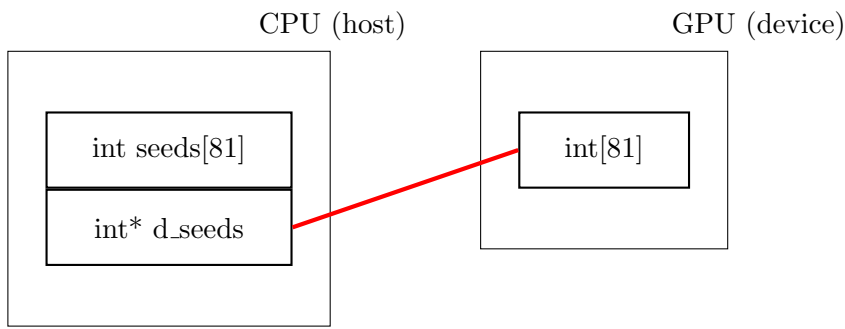

Now, at line 20 of Listing \ref{list:ex1} we allocate a memory block on the device (the GPU). The second parameter in the \textbf{cudaMalloc} command defines the total size (in bytes) of the memory block we want to allocate, i.e. \textbf{size*threads}, while the first parameter is the pointer to the allocated memory block. More information regarding the CUDA memory management could be found in \cite{cudamemory}. Figure \ref{fig:Ex1P4} corresponds to the current memory snapshot of the host and the device. We must take care not to overwrite the value of d\_seeds. Otherwise the address of the allocated device memory block is lost. Therefore, we highlighted this interconnection as a red thick line.

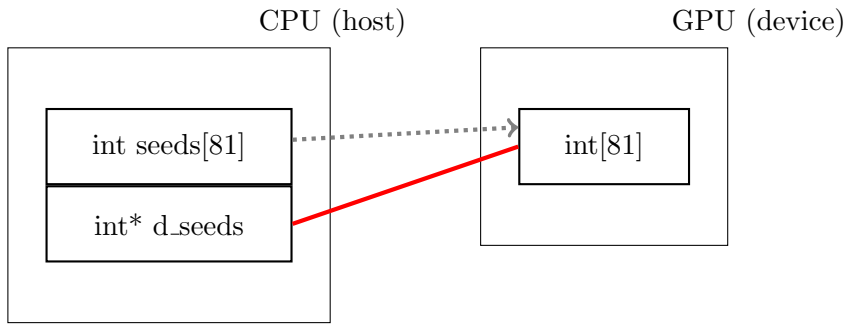
\begin{figure}[H]
\centering
\begin{tikzpicture}
  \node[block2] (ha) {int seeds[81]};
  \node[block2, below=0cm of ha] (hb) {int* d\_seeds};
  \node[block,right=3cm of ha] (da) {int[81]};
  \node[draw,inner xsep=5mm,inner ysep=8mm,fit=(ha) (hb),label={60:CPU (host)}](f){};
  \node[draw,inner xsep=5mm,inner ysep=8mm,fit=(da),label={60:GPU (device)}]{};
   \draw[red, ultra thick, -] (hb.east) -- (da.west);
  \draw[gray, ultra thick, ->, dotted] (ha)-- ($(da.south west)!0.75!(da.north west)$);
\end{tikzpicture}
\caption{cudaMemcpy() transfers the data from the host seeds array to the device d\_seeds array.}
\label{fig:Ex1P5}
\end{figure}

To proceed, we need to transfer the host memory block to the device memory block. For this purpose, we are going to use the aforementioned \textbf{cudaMemcpy} function. Line 23 of Listing \ref{list:ex1} copies the contents from the host memory block \textbf{seeds} to the device memory block \textbf{d\_seeds}. The total size of data to be transferred is defined via \textbf{size*threads}. The current snapshot of host and device memory blocks is given in Figure \ref{fig:Ex1P5}.
The gray dotted line highlights the direction of the transferred data. In this case, since we call the cudaMemcpy routine with the \enquote{cudaMemcpyHostToDevice} option, the dotted arrow is pointing to the device memory block. 

Line 25 of Listing \ref{list:ex1} calls the device function \textbf{H} 81 times in a loop from 1 to 81). So  a total of 81 threads are defined, each with an instance of H.

\begin{figure}[H]
\centering
\begin{tikzpicture}
  \node[block2] (ha) {int seeds[81]};
  \node[block2, below=0cm of ha] (hb) {int* d\_seeds};
  \node[block,right=3cm of ha] (da) {int[81]};
  \node[block,below= 0cm of da] (db) {\color{red}{Launch H}};
  \node[draw,inner xsep=5mm,inner ysep=8mm,fit=(ha) (hb),label={60:CPU (host)}](f){};
  \node[draw,inner xsep=5mm,inner ysep=8mm,fit=(da) (db),label={60:GPU (device)}]{};
   \draw[red, ultra thick, -] (hb.east) -- (da.west);
\end{tikzpicture}
\caption{After all information has been transmitted to the GPU memory, each GPU thread launches it's own instance the kernel function H.}
\label{fig:Ex1P6}
\end{figure}
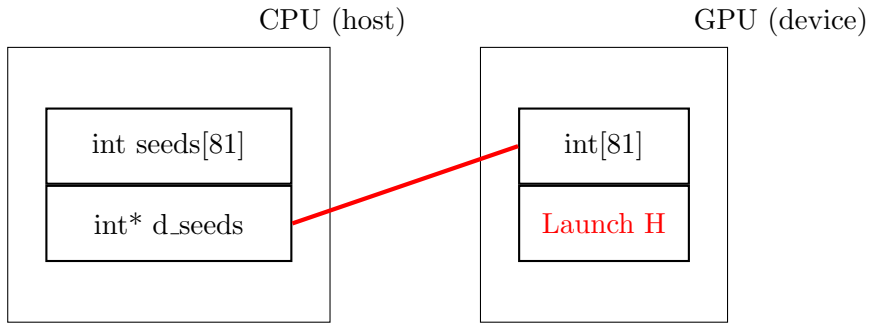

Before launching the kernel function H all required data should have been transmitted to the GPU memory. Each thread from the GPU device is going to read the corresponding memory cell from the device memory block, which is defined as the \textbf{global memory}. More details regarding the CUDA memory model are presented in Section \ref{sec:CUDADebug}. Then, the result is printed in the console output. The current snapshot is shown in Figure \ref{fig:Ex1P6}.

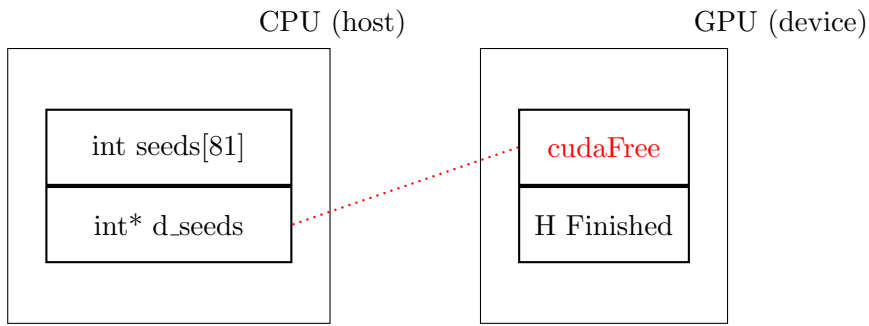
\begin{figure}[H]
\centering
\begin{tikzpicture}
  \node[block2] (ha) {int seeds[81]};
  \node[block2, below=0cm of ha] (hb) {int* d\_seeds};
  \node[block,right=3cm of ha] (da) {\color{red}{cudaFree}};
  \node[block,below= 0cm of da] (db) {H Finished};
  \node[draw,inner xsep=5mm,inner ysep=8mm,fit=(ha) (hb),label={60:CPU (host)}](f){};
  \node[draw,inner xsep=5mm,inner ysep=8mm,fit=(da) (db),label={60:GPU (device)}]{};
   \draw[red, thick, dotted, -] (hb.east) -- (da.west);
\end{tikzpicture}
\caption{Finally, when all of the 81 instances of H are finished, the memory block allocated on the device is freed.}
\label{fig:Ex1P7}
\end{figure}

Finally, \textbf{cudaFree()} frees the allocated device memory block (line 27 of Listing \ref{list:ex1}). The final snapshot of host and device is given in Figure \ref{fig:Ex1P7}.  Now the variable d\_seeds points to an un-allocated memory block in the device, which is depicted by a dotted red line.

Launching the kernel H activates all the defined threads. However, we should be very carefully in writing the kernel routines. For example, fragments of memory blocks, to which two or more threads have unintentional write access is an undesirable behavior in our program.   Let's take two different threads $t_1$ and $t_2$, and some bytes located on the device memory block denoted as $B_x$, to which both threads $t_1$ and $t_2$ have write access. While the kernel routine $t_1$ writes to $B_x$ the value $v_1$, $t_2$ writes the value $v_2$ to $B_x$. In this scenario, and when the kernel routines are finalized, we can only speculate what is the final value of $B_x$ -- $v_1$ or $v_2$. Having this in mind, we should always check for possible thread interference or deadlocks.

\begin{figure}[h]
\centering
\includegraphics[width=250px]{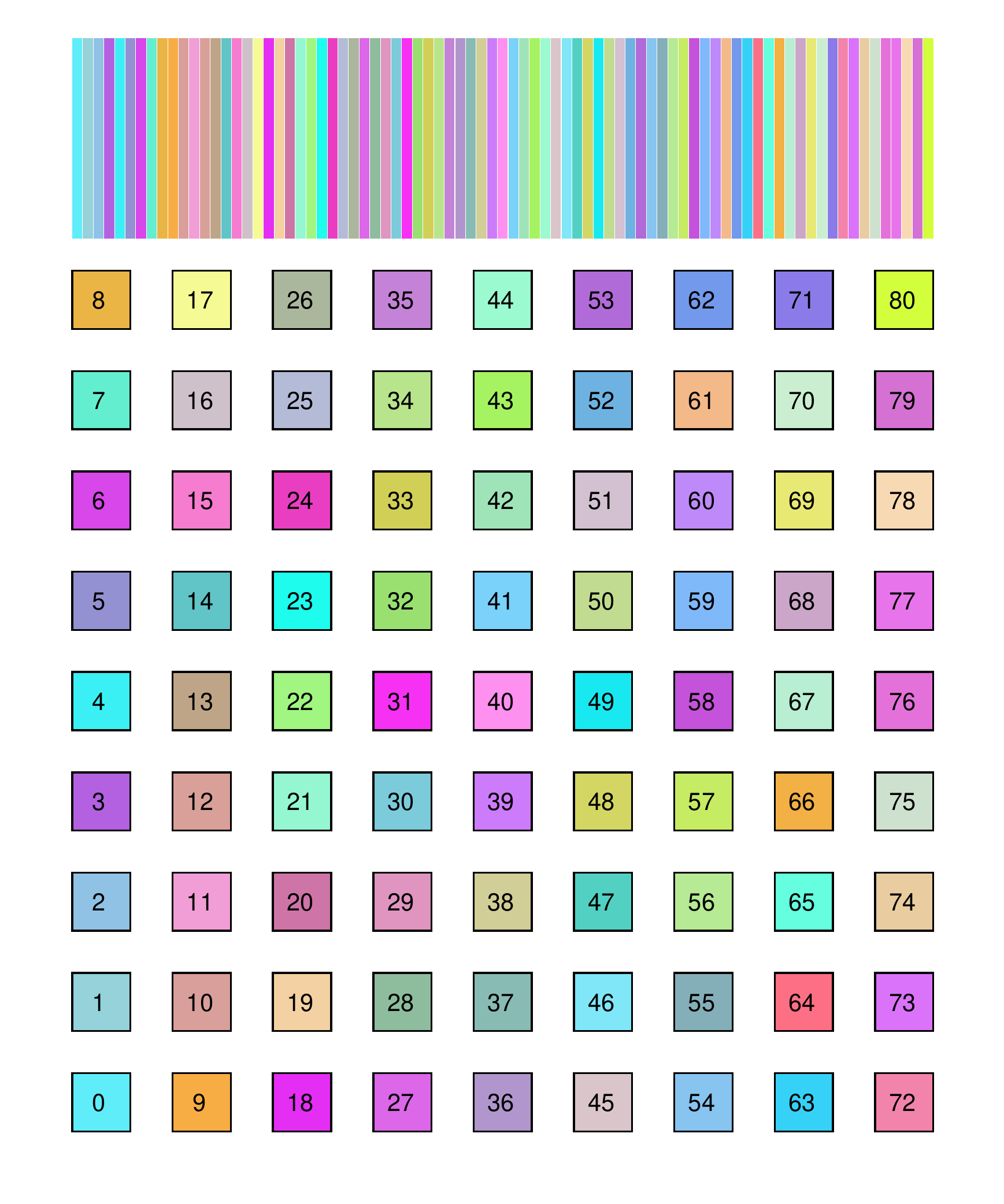} 
\caption{Visualization of the GPU global memory as a multi-color rectangle (to be found above the thread matrix). Each GPU thread and each stripe is depicted with a distinct color.}
\label{fig:ExampleMemory} 
\end{figure}

The threads on the device and the used GPU memory are visualized in Figure \ref{fig:ExampleMemory}. Since each thread has a unique thread id (via the call in Line 25 of Listing \ref{list:ex1}), each thread, labeled with an id number, is visualized as a distinct color. Above the labeled threads, an illustrative example of the GPU global memory is shown. Starting from left to right, each color stripe corresponds to the thread id sharing the same color. In this specific example, we have organized the corresponding color map from lower id numbers to higher ones, i.e. thread id 0 is at the left-most positioned stripe inside the global device memory, while thread id 80 corresponds to the final right-most stripe inside the global memory.  Each stripe is strictly related to the thread block sharing the same color.

In Listing \ref{list:outputEx4} an example output on the host terminal is given. Some lines of the output are omitted. Since each thread is independent from the others and the order of thread completion is hard to predict, the results could be reported back in a scrambled way. 

\begin{minipage}{\linewidth}
\begin{lstlisting}[label={list:outputEx4}, caption=Example output of Listing \ref{list:ex1}, style=customconsole]
console:nvcc 4Hello.cu -o hello
console:./hello
64 squared is: 4096 
65 squared is: 4225 
66 squared is: 4356 
...omitted...
78 squared is: 6084 
79 squared is: 6241 
80 squared is: 6400 
0 squared is: 0 
1 squared is: 1 
2 squared is: 4 
...omitted...
61 squared is: 3721 
62 squared is: 3844 
63 squared is: 3969 
console:
\end{lstlisting}
\end{minipage}

\subsection{Parallel squaring -- Part II}
\label{subsec:sqaring_filling_and_reading_kernel_memory}

Now, let's try another strategy -- instead of printing out the squared numbers by the kernel threads, we can copy the results calculated in the device memory block  back to the given host memory block (so the host prints the results). In fact, calling the printing function inside the kernel is not a common routine and should be avoided. The modified program is shown in Listing \ref{list:ex2}.

\begin{minipage}{\linewidth}
\begin{lstlisting}[label={list:ex2}, caption=CUDA parallel squaring with filling and reading the kernel memory from the host (two-way), style=customc, numbers=left,stepnumber=1]
#include <stdio.h>
#define threads 81

__global__ void K(int *d_seeds){
	d_seeds[threadIdx.x]=d_seeds[threadIdx.x]*d_seeds[threadIdx.x];
}

int main() {

	int seeds[threads];
	
	for (int j=0; j<threads; ++j)
		seeds[j] = j;
		
	int size = sizeof(int);
	int *d_seeds;
	
	// allocate memory block on the device
	cudaMalloc((void **)&d_seeds, size*threads);	
	
	// copy the seeds array to the device memory
	cudaMemcpy(d_seeds,seeds,size*threads,cudaMemcpyHostToDevice);
	
	K<<<1,threads>>>(d_seeds);
	cudaDeviceSynchronize();
	
	// copy the results array back to the host memory
	cudaMemcpy(seeds,d_seeds,size*threads,cudaMemcpyDeviceToHost);
	cudaFree(d_seeds);
	
	for (int j=0; j<threads; ++j)
		printf("Cell %d is equal to: %d\n", j, seeds[j]);
	
	return 0;
	
}
\end{lstlisting}
\end{minipage}

\begin{figure}[H]
\centering
\begin{tikzpicture}
  \node[block2] (ha) {int seeds[81]};
  \node[block2, below=0cm of ha] (hb) {int* d\_seeds};
  \node[block,right=3cm of ha] (da) {int[81]};
  \node[block,below= 0cm of da] (db) {H Finished};
  \node[draw,inner xsep=5mm,inner ysep=8mm,fit=(ha) (hb),label={60:CPU (host)}](f){};
  \node[draw,inner xsep=5mm,inner ysep=8mm,fit=(da) (db)
  ,label={60:GPU (device)}]{};
   \draw[gray, ultra thick, <-, dotted] (ha)-- (da);
   \draw[red, ultra thick, -] (hb) -- (da);
\end{tikzpicture}
\caption{cudaMemcpy() transfers data back from the device memory array to the host seeds array.}
\label{fig:Ex2P1}
\end{figure}
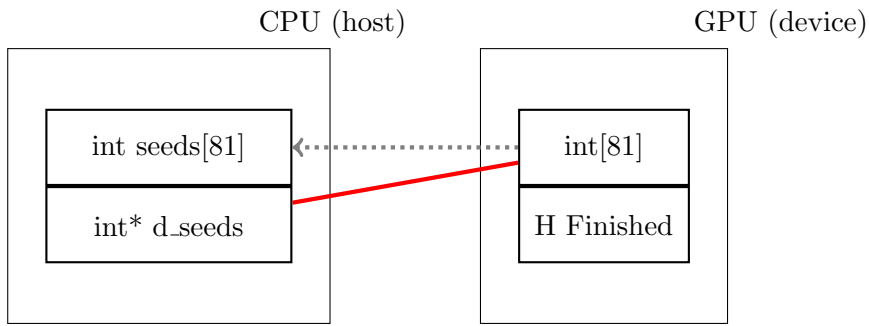

At line 5 in Listing \ref{list:ex2}, inside the newly defined kernel function \textbf{K}, we write the result of squaring the number back to the \textbf{d\_seeds} array. The main function remains almost the same as in the previous example. The only difference is that, this time (see line 28) we transfer the results back to the host memory, i.e. we copy  the data pointed by \textbf{d\_seeds} to \textbf{seeds}. The fourth argument of the \textbf{cudaMemcpy} function is \textbf{cudaMemcpyDeviceToHost}. Finally, as shown at lines 31 and 32, the results are printed out from the outside of the device. As a summary, the snapshot of the host and the device during these operations (line 28 from Listing \ref{list:ex2}) is given in Figure \ref{fig:Ex2P1}. The gray dotted line highlights the direction of the transferred data. In this case, since we call the cudaMemcpy routine with the \enquote{cudaMemcpyHostToHost} option, the dotted arrow is pointing to the host memory block. 

We have developed a handy pen-and-paper method to proactively design and architect a GPU implementation.
Now, we have all the necessary tools to complete our first project -- a full GPU-based automatic cryptanalysis tool for messages encrypted by a monoalphabetic substitution cipher.

\section{GPU cryptanalysis}
\label{sec:GPUSubstAnalysis}

In this section, we are going to launch an automatic GPU-based cryptanalysis on English messages encrypted by a substitution cipher. However, the method is universal for any language, as long as you possess the necessary statistics.

For example, in this section, we are going to exploit the partially-predictable structure of the English language by using bigrams.

\subsection{Frequency analysis}
\label{subsec:frequencyAnalysis}

Let's assume that we have a single file describing the properties of the English language bigrams. At each row of this file a single bigram with its corresponding score value is given. Since we have a total of 26 letters in the English alphabet we have a total of $26*26 = 676$ distinct rows (bigrams from \textbf{AA} to \textbf{ZZ}). Each score for a given bigram reflects the likelihood of this specific bigram to participate in an authentic English text. In short, the higher the overall score, the closer to a grammatically correct English the text is. 

For example, one of the most frequently seen bigrams in English texts are \textbf{th}, \textbf{he}, \textbf{in}, \textbf{er}, \textbf{an}, \textbf{re}, \textbf{es}, \textbf{on} and \textbf{st} (ordered by their magnitude of appearance). The likelihood of a given bigram to participate in some English text is calculated by analyzing big databases of English corpus. 

Imagine we have a huge digital library of English books. Then, by ignoring all the special symbols and by treating all the letters as lowered-cased ones, we can iterate over all the parsed texts and count the total number of appearances of distinct bigrams. For example, let's analyze the amusing quote linked to \textbf{Margaret Mead}:

\begin{displayquote}
Always remember that you are absolutely unique. \\ Just like everyone else...
\end{displayquote}

Now, we parse the text to clean it up from special symbols and to further lowercase all the letters:

\begin{displayquote}
alwaysrememberthatyouareabsolutelyuniquejustlikeeveryoneelse
\end{displayquote}

We ended up with a string of length 60, i.e. it consists of 59 bigrams we need to analyze:

\[ \text{al}, \text{lw}, \text{wa}, \cdots, \text{ls}, \text{se} \]

This resulted in the following statistics: 'al': 1, 'lw': 1, 'wa': 1, 'ay': 1, 'ys': 1, 'sr': 1, 're': 2, 'em': 2, 'me': 1, 'mb': 1, 'be': 1, 'er': 2, 'rt': 1, 'th': 1, 'ha': 1, 'at': 1, 'ty': 1, 'yo': 2, 'ou': 1, 'ua': 1, 'ar': 1, 'ea': 1, 'ab': 1, 'bs': 1, 'so': 1, 'ol': 1, 'lu': 1, 'ut': 1, 'te': 1, '\textbf{el}': 2, 'ly': 1, 'yu': 1, 'un': 1, 'ni': 1, 'iq': 1, 'qu': 1, 'ue': 1, 'ej': 1, 'ju': 1, 'us': 1, 'st': 1, 'tl': 1, 'li': 1, 'ik': 1, 'ke': 1, 'ee': 2, 'ev': 1, '\textbf{ve}': 1, 'ry': 1, 'on': 1, 'ne': 1, 'ls': 1, 'se': 1.

So, the probability of some random bigram to be \textbf{ve} or \textbf{el} is respectively $\frac{1}{59}$ or $\frac{2}{59}$. However, this is far from adequate statistics -- we have used a tiny negligible fragment of English corpus. 

Let's continue with the construction of our main tool. Since working with strings is slow, we can create a map between bigrams and some integer array of length 676\footnote{The length of this array is strictly related to the alphabet length of the specific language we are going to launch the attack at. There are total of 26 letters in the English alphabet, so we have a total number of $26*26 = 676$ distinct bigrams.}, which is going to hold down all the bigram score values. 

Let's define the English alphabet as $\Omega = \lbrace A, B, C, \cdots, Z \rbrace$. We map each consequent alphabet letter from $\Omega$ to an integer value from 0 to 25, i.e. $$A \rightarrow 0,B \rightarrow 1,C \rightarrow 2,\cdots,Z \rightarrow 25.$$ 

Let's store the score values of each of the bigrams in the array $L$ of length  ${\vert \Omega \vert}^2 = 26^2 = 676$. For each of the bigrams $\overline{\mu\nu}$ we compute it's index in $L$ as $26\mu + \nu$.

For example, the score of bigram $\overline{AB}$ is mapped to the index $26*0 + 1 = 1$ of $L$, while the score of bigram $\overline{CA}$ is mapped to the index $26*2 + 0 = 52$ of $L$. The score of the last bigram over $\Omega$, i.e. $\overline{ZZ}$, is mapped to the index $26*25 + 25 = 675$ of $L$.

Now, our first task is to read and parse the file with all the English bigrams and their corresponding score values. Then, we need to organize and map the score values into an array following the aforementioned mapping. Finally, we are going to transfer the created mapped array to the device global memory, making the array accessible within each GPU thread. The code fragment is given in Listing \ref{list:extract}.

\begin{minipage}{\linewidth}
\begin{lstlisting}[label={list:extract}, caption=Helper function to read and parse from bigram collection file, style=customc, numbers=left,stepnumber=1]
// a host function to extract and map 
// all the bigrams from a given file
__host__ void extractBigrams(unsigned long long int *scores) {
	FILE* bigramsFile = fopen("bigramParsed", "r");
	while(1){
		char tempBigram[2];
		unsigned long long int tempBigramScore = 0;
		if (fscanf(bigramsFile, "%s %llu", tempBigram, &tempBigramScore) < 2)
			 { break; } 
		scores[(tempBigram[0]-'a')*26 + tempBigram[1]-'a'] = tempBigramScore; 
	}
	fclose(bigramsFile);
}
\end{lstlisting}
\end{minipage}

As shown in the previous source code fragment, we named the bigram collection file as \textbf{bigramParsed} (line 4). Then, by reading the file line by line, we repeatedly extract each bigram and then compute the corresponding score (see line 8). We recall that the \textbf{fscanf} function reads data from the current position of the specified stream into the locations that are given by the entries in the argument-list. The actual mapping is done at line 10. To achieve that, we subtract the ascii value of \textbf{a} from each letter from the bigram, so:

\begin{itemize}
\item{\textbf{a} is mapped to \textbf{0}}
\item{\textbf{b} is mapped to \textbf{1}} 
\item{$\cdots$}
\item{\textbf{z} is mapped to \textbf{25}}
\end{itemize}

Having this in mind, the helper function is adequate to low-cased bigrams collection related files only. If you want to use another bigram format then a further modification of the function logic is needed. 

Finally, we save all the scores to an \textbf{unsigned long long int}  array named \textbf{scores}. The main function of our GPU cryptanalysis tool is given in listing \ref{list:ex3}. Take a brief look, but don't try to grasp all the details and \enquote{magic} constants yet. A detailed explanation accompanied by helpful comments is given right after the source code fragment.

\begin{lstlisting}[label={list:ex3}, caption=The main function of the CUDA \ac{mas} cryptanalysis project, style=customc, numbers=left,stepnumber=1]
int main() {

	srand(time(0));
	int totalBigrams = 676;
	int threads = 325;
	char encrypted[] = <omitted>;
	int encryptedLen = strlen(encrypted);
	int encryptedMap[encryptedLen];
	
	unsigned long long int results[threads];
	
	// we map the encrypted message to integers
	for (int j=0; j<encryptedLen; ++j)
		encryptedMap[j] = encrypted[j] - 'a';
	
	unsigned long long int scores[totalBigrams];	
	extractBigrams(scores);
		
	int size = sizeof(unsigned long long int);
	unsigned long long int *d_scores;
	unsigned long long int *d_results;
	int *d_encrypted;
	
	// allocate memory blocks on the device
	cudaMalloc((void **)&d_scores, size*totalBigrams);
	cudaMalloc((void **)&d_results, size*threads);
	cudaMalloc((void **)&d_encrypted, sizeof(int)*encryptedLen);	
	
	// copy the scores array to the device memory (only once)
	cudaMemcpy(d_scores, scores, size*totalBigrams, cudaMemcpyHostToDevice);
	
	int leftLetter = 0;
	int rightLetter = 0;
	int tempMaxIndex = 0;
	int tempMaxLeft = 0;
	int tempMaxRight = 0;
	unsigned long long int maxScore = 0;
	
	for (int j=0; j<500; ++j) 	{
	
		// pick two random distinct letters; we make sure they do exists in the encrypted text
		leftLetter = encrypted[rand() % encryptedLen] - 'a';
		rightLetter = leftLetter;
		while (rightLetter == leftLetter)
			rightLetter = encrypted[rand() % encryptedLen] - 'a';
	
		// copy the encrypted msg array to the device memory (after each climbing)
		cudaMemcpy(d_encrypted, encryptedMap, sizeof(int)*encryptedLen, cudaMemcpyHostToDevice);
	
		K<<<1,threads>>>(d_scores, d_encrypted, encryptedLen, leftLetter, rightLetter, d_results);
		cudaDeviceSynchronize();
	
		// copy the results array back to the host memory
		cudaMemcpy(results, d_results, size*threads, cudaMemcpyDeviceToHost);
	
		tempMaxIndex = getMaxElement(results, threads);
		
		if (results[tempMaxIndex] > maxScore) {
			maxScore = results[tempMaxIndex];
			tempMaxLeft = 0;
			tempMaxRight = 0;
			translateIndexToLetters(tempMaxIndex, &tempMaxLeft, &tempMaxRight);
			climb(encryptedMap, encryptedLen, leftLetter, tempMaxLeft, rightLetter, tempMaxRight);
			printf("%d %llu\n", tempMaxIndex, results[tempMaxIndex]);
			printf("%d %d %d %d\n", tempMaxLeft, tempMaxRight, leftLetter, rightLetter);
			demap(encryptedMap, encryptedLen);
			
		}
	}
	
	// we free the allocated device memory 
	cudaFree(d_scores);
	cudaFree(d_encrypted);
	cudaFree(d_results);
	
	return 0;	
}
\end{lstlisting}

Line 3 initializes the seed of the \ac{prng}, which is going to be repeatedly utilized throughout the main function core routine. There are a total of $26^2 = 676$ bigrams over the English language (line 4). The encrypted message is defined as a string with name \textbf{encrypted} (line 6). We further take its length (line 7), and make a declaration of another array of integers named \textbf{encryptedMap}, of the same length as the encrypted message itself (line 8). 

\subsection{On the multi-threaded hill climbing design}

Now let's pause for a while. Since we want to utilize the multi-core capabilities of the GPU, we are going to perform a \textbf{multi-threaded hill-climbing} method to automatically decrypt a given message encrypted by \ac{mas} cipher. Hill climbing is a mathematical optimization technique and an iterative algorithm that starts with an arbitrary solution to a problem, then attempts to find a better solution by making an incremental change to the solution. If the change produces a better solution, another incremental change is made to the new solution, and so on until no further improvements can be found. More details can be found in \cite{lasry2018methodology}.

Now, for simplicity, let's denote the encrypted message as $E$, which was constructed over some alphabet $\Omega$. 

Let's denote the key of the \ac{mas} cipher as $K$. We can decompose $K$ to 26 single substitutions, i.e.:
$$a \rightarrow \zeta_0$$
$$b \rightarrow \zeta_1$$
$$\cdots$$
$$z \rightarrow \zeta_{25,}$$ where $\zeta_i \in \Omega$ and $\forall i,j \implies \zeta_i \neq \zeta_j$. A \textbf{normal hill-climbing} approach to recover an unknown key of such encryption scheme can be summarized with the following   steps:

\begin{enumerate}
\item{Generate two random letters $\psi_1$ and $\psi_2$ over the alphabet $\Omega$.}
\item{Given an encrypted message $E$, interchange all the letters $\psi_1$ with $\psi_2$. Let's denote the resulted message as $E'$.}
\item{By using bigrams (or any other statistics) compare the scores of $E$ and $E'$. If $E'$ is a better candidate than $E$, we overwrite $E$ with $E'$.}
\item{Repeat until some threshold value is reached.}
\end{enumerate}

Since we want to benefit from the \textbf{multi-core capabilities} of the GPU, we can tweak and modify the normal hill climbing algorithm with the following multi-threaded hill variant:

\begin{enumerate}
\item{Generate two random distinct letters $\psi_l$ and $\psi_r$ over the alphabet $\Omega$. Given an encrypted message $E$, we make sure that both $\psi_l,\psi_r \in E$.}  

\item{Launch the kernel function in such a way, that every distinct thread corresponds to a unique unordered pair of distinct letters $\delta_l$ and $\delta_r$ over the alphabet $\Omega$. Let us define such mapping as $\Upsilon$.}

Example: Let's say we have an encrypted message $E$ over some reduced alphabet consisting of the letters \textbf{a}, \textbf{b} and \textbf{c} only. Since the reduced alphabet consists of 3 letters, the total count of unique pairs of distinct letters $\delta_l$ and $\delta_r$ over $\Omega$ is $\frac{3*2}{2} = 3$. Indeed, we have the following possible set of ordered pairs $\delta_l$ and $\delta_r$:

\[ \left( \delta_l, \delta_r \right) = \left\lbrace \left( a, b \right), \left( a, c \right), \left( b, a \right), \left( b, c \right), \left( c, a \right), \left( c, b \right)  \right\rbrace \]

.. which is reduced to the following set of unordered pairs $\delta_l$ and $\delta_r$:

\[ \left( \delta_l, \delta_r \right) = \left\lbrace \left( a, b \right), \left( a, c \right), \left( b, c \right) \right\rbrace \]

Having this in mind, we need to utilize a total number of 3 GPU threads. Let's denote them as $t_1, t_2$ and $t_3$. A possible mapping $\Upsilon_1$ could be:

\[ \Upsilon_1: t_1 \rightarrow (b,c); t_2 \rightarrow (a,c); t_3 \rightarrow (a,b) \]

This is a valid mapping, since each thread is mapped to a distinct pair and there is a total number of 3 mappings.    

\item{Given an encrypted message $E$, each thread should interchange $\psi_l$ with $\delta_l$, as well as $\psi_r$ with $\delta_r$. However, to guarantee that the order of the interchanges is irrelevant, we make sure that they take place  if and only if $\psi_l \neq \delta_r$ and $\psi_r \neq \delta_l$. Let's denote the resulted message as $E'$.}

\item{By using bigrams, each distinct thread calculates the corresponding score of $E'$ denoted by $F(E')$. Then, the score $F(E')$ is further saved to the corresponding thread memory cell. When all the threads are ready and synchronized, we transfer back all the scores to the host.}

\item{We analyze all the scores and pick up the best yielded score. Let's denote the best score on index $i$ as $F(E^{'}_{i})$. If $E^{'}_{i}$ possesses a better score than $E$, we overwrite $E$  with $E'_{i}$.

We should emphasize that currently we have $F(E^{'}_{i})$, not $E^{'}_{i}$ itself. However, by taking the index of the best yielded score, i.e. $i$, we can reconstruct the actual value of $E^{'}_{i}$. We can easily achieve that by applying the inverse mapping $\Upsilon^{-1}$ on thread number $i$.}   

Example: We use the same scenario as in step 2. We have the following setup before launching the GPU threads on the encrypted message $E$:

\[ \Upsilon_1: t_1 \rightarrow (b,c); t_2 \rightarrow (a,c); t_3 \rightarrow (a,b) \]

The  letters  $\psi_l,\psi_r \in E$ are chosen pseudo-randomly. By applying the interchanges described above, a given text $E_i$ is computed by the corresponding thread $t_i$, i.e. $t_1 \rightarrow E_1, t_2 \rightarrow E_2, $ and $t_3 \rightarrow E_3$. Then, the concatenated values of the scores of $E_i$ are transferred back to the host. Let's store these values in the array $F(E_1)|F(E_2)|F(E_3)$. Now, suppose that $F(E_2)$ is the best among all the scores. We further recall that the host doesn't have the $E_2$ text itself at the moment. However, we know the position of $F(E_2)$, i.e. $2$, which corresponds to the index of the thread that yielded this score, e.g. $t_2$. By using the inversion mapping of $\Upsilon_1$:

\[ \Upsilon_1^{-1}: (b,c) \rightarrow t_1; (a,c)  \rightarrow t_2; (a,b) \rightarrow  t_3, \]

we can trace-back the operation on $E$, which yielded $E_2$ and, therefore, fully recover it.

\item{We repeat until some threshold value is reached. This value usually depends on the size of the set of the possible initial choices (in our implementation we have two choices -- $\psi_l$ and $\psi_r$).}
\end{enumerate}

One could ask -- "What is the reason to introduce step 1? Why don't we just try all the possible unordered pairs $(\delta_i, \delta_j)$ inside the GPU?". Well, such strategy would benefit from the multi-core architecture of the GPU as well. However, there is one major drawback -- it is deterministic. To illustrate that, let's start from some encrypted message $E$ over some alphabet $\Omega$, s.t. $\vert\Omega\vert = n$. We transfer $E$ to the GPU, together with all the necessary data to evaluate the generated candidates. We skip step 1 and we do not generate two letters $\psi_i$ and $\psi_j$, s.t. $\psi_i, \psi_j \in \Omega$. Each GPU thread corresponds to a given transposition of letters in $E$. Without loss of generality, we can represent the thread mapping in lexicographical order, i.e.:

\begin{align*}
 &\Upsilon: \\
 &t_1 \rightarrow (\delta_1, \delta_2)  \\ 
 &t_2 \rightarrow (\delta_1, \delta_3) \\
 &\cdots  \\ 
 &t_{x} \rightarrow (\delta_i, \delta_j), i<j \\
 &\cdots \\
 &t_{\frac{n(n-1)}{2}} \rightarrow (\delta_{n-1}, \delta_{n})
\end{align*}

Now, let's denote the generated candidate from a given thread $t_i$ as $E_i$. Furthermore, let's denote the score results of $E_i$ under some evaluating function $F$ as $F(E_i)$. For simplicity, we choose $F$ in such a way, that  \[ \forall i,j \in \left[1,T\right], i \neq j : F(E_i) \neq F(E_j). \] with $T = \frac{n(n-1)}{2}$. Then, we have the following composition:

\begin{align*}
 &\Upsilon \circ F: \\
 &t_1 \rightarrow (\delta_1, \delta_2) \rightarrow E_1 \rightarrow F(E_1) \\ 
 &t_2 \rightarrow (\delta_1, \delta_3) \rightarrow E_2 \rightarrow F(E_2) \\
 &\cdots  \\ 
 &t_{x} \rightarrow (\delta_i, \delta_j) \rightarrow E_x \rightarrow F(E_x) \\
 &\cdots \\
 &t_T \rightarrow (\delta_{n-1}, \delta_{n}) \rightarrow E_T \rightarrow F(E_T)
\end{align*}

Then, we collect all the resulting scores and choose the best candidate $E_m$, s.t. \[ \forall i \in \left[1,T\right], i \neq m : F(E_i) < F(E_m). \] 

Let's denote as $M_1$ the value of the maximum achieved score, i.e. \[ M_1 = \max_{i=1}^{T}{F(E_i)} = F(E_m)\]

Now, if we restart the process from the beginning, the best yielded score will be exactly the same value, i.e. $M_1$. However, if we do not restart the process the optimization process will proceed as usual -- since $F(E_m)$ can be found at the position $m$, then the thread $m$ yielded this score, i.e. $t_m$. Hence, we interchange the current message $E$ with the reconstructed better candidate $E_m$. Then, we repeat the procedure until we have reached a local maximum candidate $E_{max}$. Let's denote  the best score yielded on iteration $i$ as $M_i$. The history of our best yielded scores during the optimization process could be summarized as follows:

\[ M_1, M_2, \cdots, M_{k-1}, M_k, \text{NA} \]

where \textbf{NA} is the abbreviation of "not available", denoting the absence of a better score after the k-th iteration step. 

We can generalize our previous observation, by stating that the restart of the optimization process from a given encrypted message $E_x$, which corresponds to, for example, a score $M_x$, and applying only one iteration, will lead to the best yielded score $M_{x+1}$. Therefore, if we restart the process from the beginning, we will reach $M_1$, then $M_2$, $\cdots$, then $M_k$, to travel exactly the same optimization path, and to reach exactly the same final local optimum. In this scenario, having a complete deterministic metaheuristic, we define $M_k$ as the attractor of $M_1$. 

\subsection{Implementation}
\label{subsec:Implement}

Now, in the context of our current problem, we can define the bijective mapping $\Upsilon$ as a lexicographical order mapping. Since we have a total of 26 letters, without considering the interchanging restriction introduced in step 3, we need a total of ${26 \choose 2} = 325$ threads (see line 5 of the \textbf{main} function source code).

\begin{figure}[h]
\centering
\begin{tikzpicture}
  \node[block3] (ha) {encrypted[encryptedLen]};
  \node[block3, below=0cm of ha] (hb) {encryptedMap[encryptedLen]};
  \node[block3, below=0cm of hb] (hc) {results[threads]};
  \node[block3, below=0cm of hc] (hd) {scores[totalBigrams]};
  \node[block,right=3cm of ha] (da) {empty};
  \node[draw,inner xsep=5mm,inner ysep=8mm,fit=(ha) (hb) (hc) (hd),label={60:CPU (host)}](f){};
  \node[draw,inner xsep=5mm,inner ysep=8mm,fit=(da)
  ,label={60:GPU (device)}]{};
\end{tikzpicture}
\caption{First steps of initializing the algorithm presented in Listing \ref{list:ex3}.}
\label{fig:Ex3P1}
\end{figure}
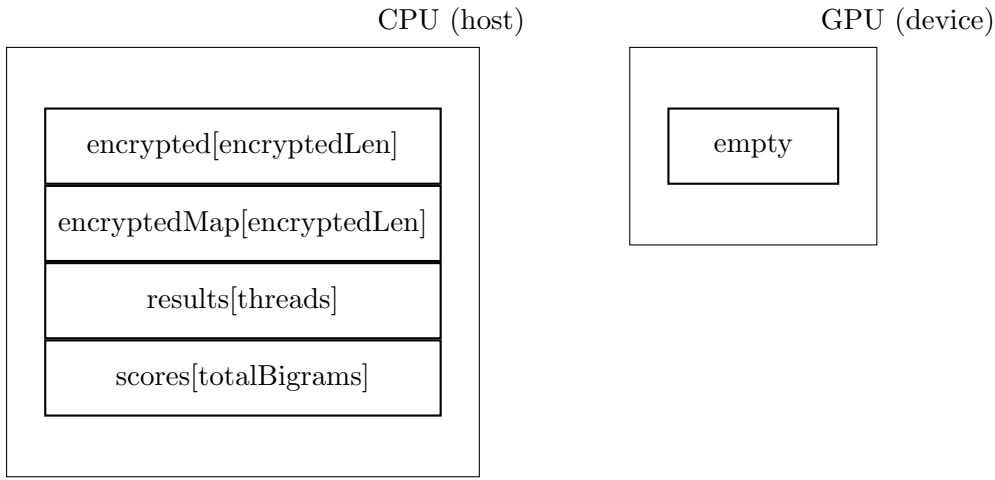

We save all the scores reported back from the device threads into the \textbf{results} array (defined at line 10). A regular mapping from ASCII to integer indexes is performed on the encrypted text as well (lines 12-14). The current snapshot of the memory blocks of the host and the device is given in Figure \ref{fig:Ex3P1}.

At the beginning, we read the encrypted message from a given file and save it in the character array \textbf{encrypted}. Then, we translate the character message to an integer one by using the integer array \textbf{encryptedMap}. We continue by creating an array \textbf{results} which is going to keep the scores yielded by each GPU thread. Finally, we populate an array named \textbf{scores} with the bigram scores we have prepared (lines 16 and 17 in Listing \ref{list:ex3}).

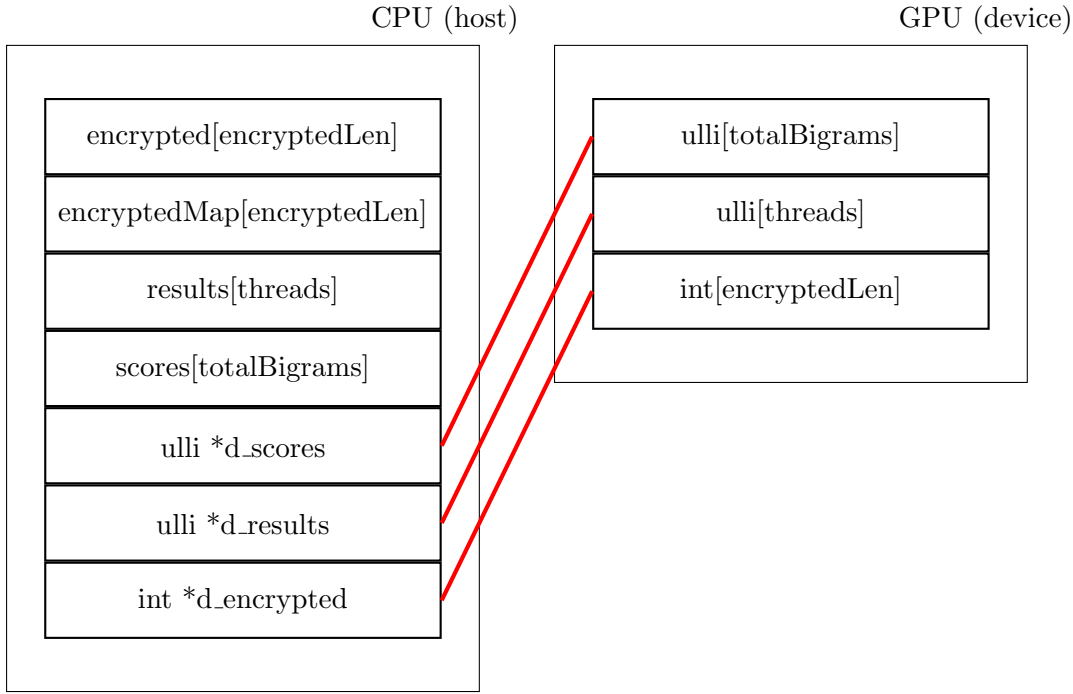
\begin{figure}[H]
\centering
\begin{tikzpicture}
  \node[block3] (ha) {encrypted[encryptedLen]};
  \node[block3, below=0cm of ha] (hb) {encryptedMap[encryptedLen]};
  \node[block3, below=0cm of hb] (hc) {results[threads]};
  \node[block3, below=0cm of hc] (hd) {scores[totalBigrams]};
  \node[block3, below=0cm of hd] (he) {ulli *d\_scores};
  \node[block3, below=0cm of he] (hf) {ulli *d\_results};
  \node[block3, below=0cm of hf] (hg) {int *d\_encrypted};
  \node[block3,right=2cm of ha] (da) {ulli[totalBigrams]};
  \node[block3,below= 0cm of da] (db) {ulli[threads]};
  \node[block3,below= 0cm of db] (dc) {int[encryptedLen]};
  \node[draw,inner xsep=5mm,inner ysep=7mm,fit=(ha) (hb) (hc) (hd) (he) (hf) (hg) ,label={70:CPU (host)}](f){};
  \node[draw,inner xsep=5mm,inner ysep=7mm,fit=(da) (db) (dc)  ,label={60:GPU (device)}]{};
   \draw[red, ultra thick, -] (he.east) -- (da.west);
    \draw[red, ultra thick, -] (hf.east)-- (db.west);
     \draw[red, ultra thick, -] (hg.east)-- (dc.west);
\end{tikzpicture}
\caption{Creating the GPU memory blocks we are going to use.}
\label{fig:Ex3P2}
\end{figure}

We are ready to prepare and allocate all the necessary memory blocks inside the device global memory (lines 19-30). The memory block snapshot is depicted in Figure \ref{fig:Ex3P2}.

First, we create the necessary pointers. Then, the required space is allocated on the GPU global memory (lines 19-27 in Listing \ref{list:ex3}). For simplicity, we denote the variable type \textbf{unsigned long long int} with \textbf{ulli}. 

\begin{figure} 
\centering
\begin{tikzpicture}
  \node[block3] (ha) {encrypted[encryptedLen]};
  \node[block3, below=0cm of ha] (hb) {encryptedMap[encryptedLen]};
  \node[block3, below=0cm of hb] (hc) {results[threads]};
  \node[block3, below=0cm of hc] (hd) {scores[totalBigrams]};
  \node[block3, below=0cm of hd] (he) {ulli *d\_scores};
  \node[block3, below=0cm of he] (hf) {ulli *d\_results};
  \node[block3, below=0cm of hf] (hg) {int *d\_encrypted};
  \node[block3, below=0cm of hg] (hh) {FOR[500]};
   \node[block3, below=3cm of hh] (fa) {empty};
  \node[block3,right=2cm of ha] (da) {ulli[totalBigrams]};
  \node[block3,below= 0cm of da] (db) {ulli[threads]};
  \node[block3,below= 0cm of db] (dc) {int[encryptedLen]};
  \node[draw,inner xsep=5mm,inner ysep=7mm,fit=(ha) (hb) (hc) (hd) (he) (hf) (hg) (hh) ,label={70:CPU (host)}](f){};
   \node[draw,inner xsep=5mm,inner ysep=7mm,fit=(fa) ,label={70:FOR cycle}](f){};
  \node[draw,inner xsep=5mm,inner ysep=7mm,fit=(da) (db) (dc)  ,label={60:GPU (device)}]{};
   \draw[red, ultra thick, -] (he.east) -- (da.west);
    \draw[red, ultra thick, -] (hf.east)-- (db.west);
     \draw[red, ultra thick, -] (hg.east)-- (dc.west);
   \draw[gray, ultra thick, ->, dotted] ($(hd.south east)!0.75!(hd.north east)$) -- ($(da.south west)!0.75!(da.north west)$);
   \draw[lightgray, thick, <->, double] (hh.south)-- (fa.north);
\end{tikzpicture}
\caption{The GPU transfer of the bigram scores and launching the FOR cycle.}
\label{fig:Ex3P3}
\end{figure}
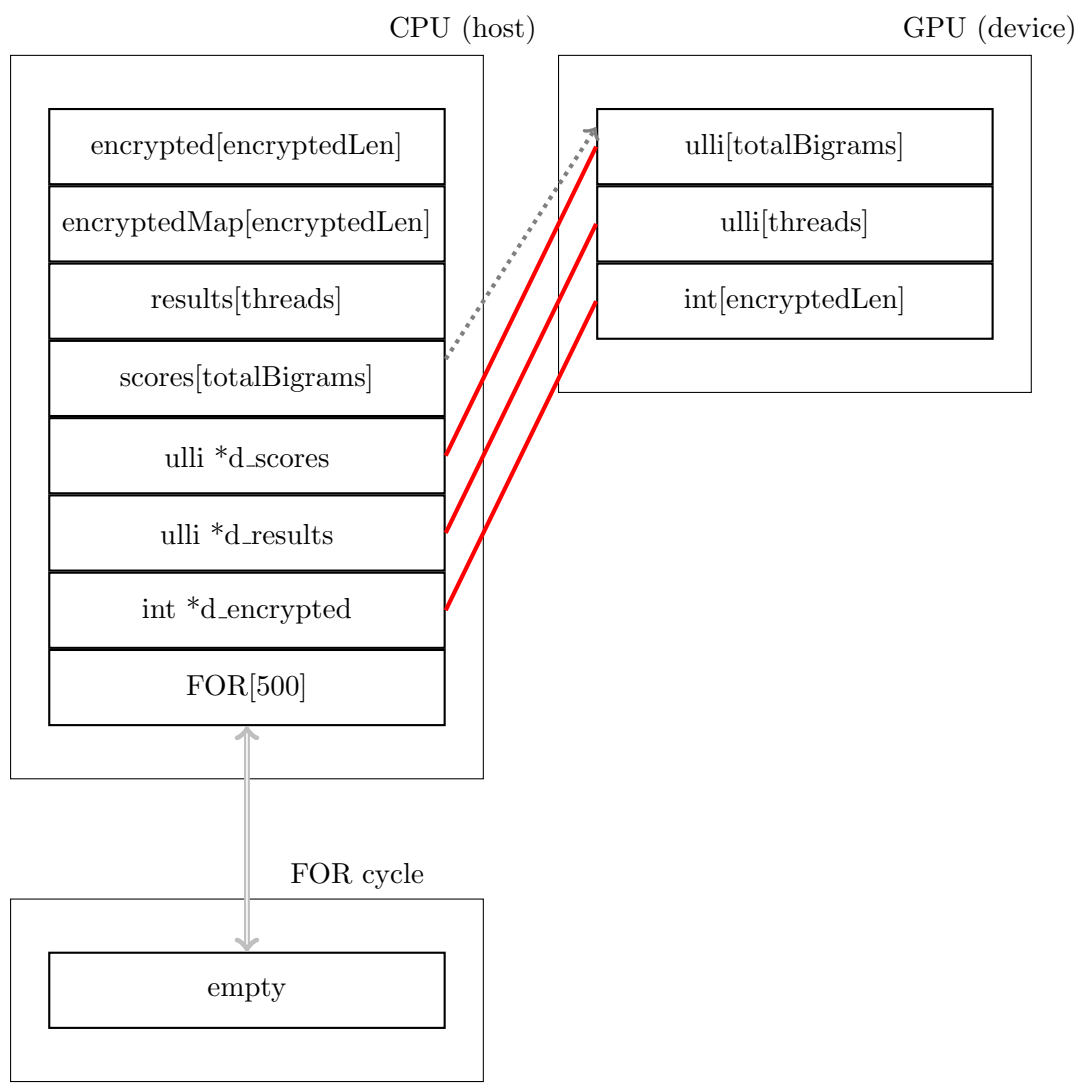

Then, the bigram scores are transferred from the host to the device. After that, we enter the \textbf{for} block. The snapshot of the memory blocks of the host and the device, right before entering the \textbf{for} cycle, is given in Figure \ref{fig:Ex3P3}.

We transfer to the GPU the bigram scores (line 30 in Listing \ref{list:ex3}). This is done just once. Next, we enter a \textbf{for} cycle, which is iterated 500 times.

Starting from line 39 to line 69, we initiate the actual multi-threaded hill climbing instance. We define a reasonable number of iterations value of 500 (at line 39). By reasonable, we are emphasizing on the negligible probability of missing a better candidate, if such exists.

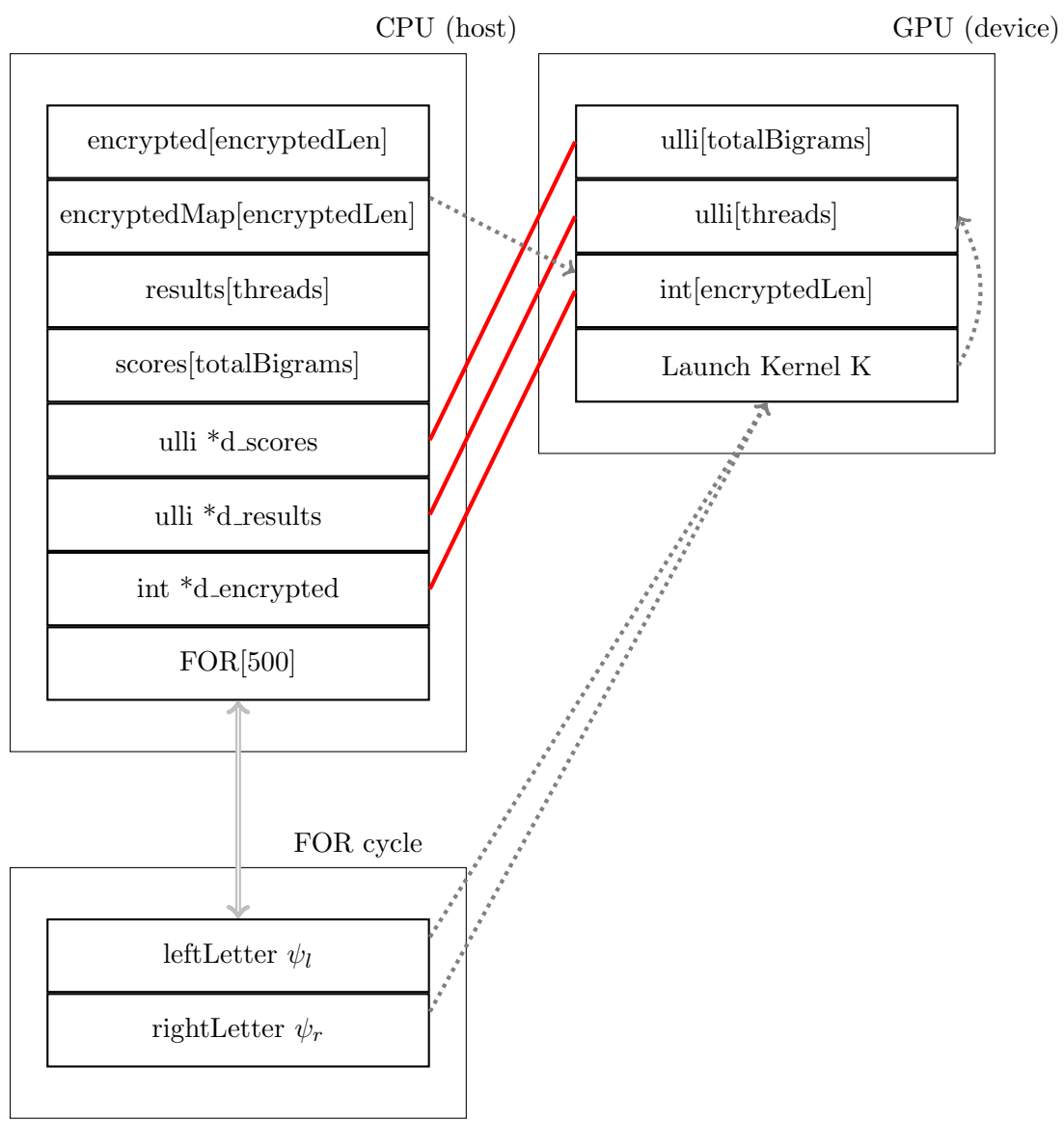
\begin{figure}
\centering
\begin{tikzpicture}
  \node[block3] (ha) {encrypted[encryptedLen]};
  \node[block3, below=0cm of ha] (hb) {encryptedMap[encryptedLen]};
  \node[block3, below=0cm of hb] (hc) {results[threads]};
  \node[block3, below=0cm of hc] (hd) {scores[totalBigrams]};
  \node[block3, below=0cm of hd] (he) {ulli *d\_scores};
  \node[block3, below=0cm of he] (hf) {ulli *d\_results};
  \node[block3, below=0cm of hf] (hg) {int *d\_encrypted};
  \node[block3, below=0cm of hg] (hh) {FOR[500]};
   \node[block3, below=3cm of hh] (fa) {leftLetter $\psi_l$};
   \node[block3, below=0cm of fa] (fb) {rightLetter $\psi_r$};
  \node[block3,right=2cm of ha] (da) {ulli[totalBigrams]};
  \node[block3,below= 0cm of da] (db) {ulli[threads]};
  \node[block3,below= 0cm of db] (dc) {int[encryptedLen]};
  \node[block3,below= 0cm of dc] (dd) {Launch Kernel K};
  \node[draw,inner xsep=5mm,inner ysep=7mm,fit=(ha) (hb) (hc) (hd) (he) (hf) (hg) (hh) ,label={70:CPU (host)}](f){};
   \node[draw,inner xsep=5mm,inner ysep=7mm,fit=(fa) (fb) ,label={70:FOR cycle}](f){};
  \node[draw,inner xsep=5mm,inner ysep=7mm,fit=(da) (db) (dc) (dd)  ,label={60:GPU (device)}]{};
   \draw[red, ultra thick, -] (he.east) -- (da.west);
    \draw[red, ultra thick, -] (hf.east)-- (db.west);
     \draw[red, ultra thick, -] (hg.east)-- (dc.west);
   \draw[gray, ultra thick, ->, dotted] ($(hb.south east)!0.75!(hb.north east)$) -- ($(dc.south west)!0.75!(dc.north west)$);
   \draw[lightgray, thick, <->, double] (hh.south)-- (fa.north);
   \draw[gray, ultra thick, ->, dotted] ($(fa.south east)!0.75!(fa.north east)$) -- (dd.south);
   \draw[gray, ultra thick, ->, dotted] ($(fb.south east)!0.75!(fb.north east)$) -- (dd.south);
   \draw[gray, bend right, ultra thick,->, dotted]  (dd.east) to [swap] (db.east);
\end{tikzpicture}
\caption{Launching kernel K.}
\label{fig:Ex3P4}
\end{figure}

First, we generate two random distinct letters $\psi_l$ and $\psi_r$ over the English alphabet, which are to be seen at least once in the encrypted text (see lines 41-45). We further transfer the encrypted message to the device -- recall that the message is not static and is always replaced in the cases when better candidate messages are found. By completing the aforementioned routines, we are ready to launch the kernel K. As soon as the kernel is activated, each thread applies the corresponding (unique) interchange $(\psi_l \rightarrow \delta_l, \psi_r \rightarrow \delta_r)$ and further calculates the score of the modified text. Then, the score is saved to the dedicated cell inside the \textbf{threads} array. Completing all the steps described in this paragraph will yield the memory block snapshot of the host and the device shown in Figure \ref{fig:Ex3P4}.

By using the host (CPU) \ac{prng} we generate two distinct letters \textbf{leftLetter} and \textbf{rightLetter} and transfer the encrypted text to the device (line 47 in Listing \ref{list:ex3}). We launch the kernel \textbf{K}. Once the routine is finished, the results are populated inside the d\_results array on the GPU.

Then, we transfer back the results (lines 48-54), to further analyze them and extract the best score. The extraction of the best score is given by the following helper function \textbf{getMaxElement} (see Listing \ref{listing:9}): 

\begin{minipage}{\linewidth}
\begin{lstlisting}[label={listing:9}, caption=The \textbf{getMaxElement} helper function, style=customc, numbers=left,stepnumber=1]
// a helper function to return the maximum element
__host__ int getMaxElement(unsigned long long int *scores, int length) {
	unsigned long long int tempMaxScore = 0;
	int tempIndex = 0;	
	for (int j=0; j<length; ++j) {
		if (scores[j] > tempMaxScore) {
			tempIndex = j;
			tempMaxScore = scores[j];
		} 
	}	
	return tempIndex;
}
\end{lstlisting}
\end{minipage}

If better candidate $E'_{i}$ is found, we accept it and update the current state (lines 58-69). Several helper functions are called inside this routine. In order to recover the candidate $E'_{i}$, as we have previously discussed, we need to calculate the $\Upsilon^{-1}$ bijective mapping. This is done by the helper function \textbf{translateIndexToLetters} (see Listing \ref{listing:10}):

\begin{minipage}{\linewidth}
\begin{lstlisting}[label={listing:10}, caption=The \textbf{translateIndexToLetters} helper function, style=customc, numbers=left,stepnumber=1]
// translate a given index to a map of pair of distinct letters 
__host__ void translateIndexToLetters(int index, int* left, int* right) {

	int tempLeft = 0;
	int tempRight = tempLeft+1;
	
	while (index > 0) {
		tempRight += 1;
		index -= 1;
		if (tempRight == 26) {
			tempLeft += 1;
			tempRight = tempLeft + 1;
		}
	}
	
	*left = tempLeft;
	*right = tempRight;
}
\end{lstlisting}
\end{minipage}

Next, we continue with the \textbf{main} function routines. We accept the better candidate by calling the helper function \textbf{climb} (see Listing \ref{listing:11}):

\begin{minipage}{\linewidth}
\begin{lstlisting}[label={listing:11}, caption=The \textbf{climb} helper function, style=customc, numbers=left,stepnumber=1]
// interchange left with maxLeft and right with maxRight
__host__ void climb(int* encrMap, int length, int left, int maxLeft, int right, int maxRight) {
	for (int j=0; j<length; ++j) {
		if (encrMap[j] == left)
			encrMap[j] = maxLeft;
		else if (encrMap[j] == maxLeft)
			encrMap[j] = left;
		else if (encrMap[j] == right)
				encrMap[j] = maxRight;
		else if (encrMap[j] == maxRight)
				encrMap[j] = right;
	}
}
\end{lstlisting}
\end{minipage}

In case a better candidate is found, we print out some details about the candidate (lines 64 and 65), as well as the current state of the message to be decrypted (line 66). This is done by the helper function \textbf{demap} (see Listing \ref{listing:12}), which should not be mistaken with the bijective mapping $\Upsilon$:

\begin{minipage}{\linewidth}
\begin{lstlisting}[label={listing:12}, caption=The \textbf{demap} helper function, style=customc, numbers=left,stepnumber=1]
// demap and print a given array of integer letters
__host__ void demap(int* encrMap, int length) {
	for (int j=0; j<length; ++j) {
		printf("%c", encrMap[j] + 'a');
	}
	printf("\n");
}
\end{lstlisting}
\end{minipage}

\begin{figure}[H]
\centering
\begin{tikzpicture}
  \node[block3] (ha) {encrypted[encryptedLen]};
  \node[block3, below=0cm of ha] (hb) {encryptedMap[encryptedLen]};
  \node[block3, below=0cm of hb] (hc) {results[threads]};
  \node[block3, below=0cm of hc] (hd) {scores[totalBigrams]};
  \node[block3, below=0cm of hd] (he) {ulli *d\_scores};
  \node[block3, below=0cm of he] (hf) {ulli *d\_results};
  \node[block3, below=0cm of hf] (hg) {int *d\_encrypted};
  \node[block3, below=0cm of hg] (hh) {FOR[500]};
   \node[block3, below=3cm of hh] (fa) {leftLetter};
   \node[block3, below=0cm of fa] (fb) {rightLetter};
  \node[block3,right=2cm of ha] (da) {ulli[totalBigrams]};
  \node[block3,below= 0cm of da] (db) {ulli[threads]};
  \node[block3,below= 0cm of db] (dc) {int[encryptedLen]};
  \node[block3,below= 0cm of dc] (dd) {kernel K};
  \node[draw,inner xsep=5mm,inner ysep=7mm,fit=(ha) (hb) (hc) (hd) (he) (hf) (hg) (hh) ,label={70:CPU (host)}](f){};
   \node[draw,inner xsep=5mm,inner ysep=7mm,fit=(fa) (fb) ,label={70:FOR cycle}](f){};
  \node[draw,inner xsep=5mm,inner ysep=7mm,fit=(da) (db) (dc) (dd)  ,label={60:GPU (device)}]{};
   \draw[red, ultra thick, -] (he.east) -- (da.west);
    \draw[red, ultra thick, -] (hf.east)-- (db.west);
     \draw[red, ultra thick, -] (hg.east)-- (dc.west);
   \draw[gray, ultra thick, <-, dotted] ($(hc.south east)!0.75!(hc.north east)$) --  ($(db.south west)!0.75!(db.north west)$) ;
   \draw[lightgray, thick, <->, double] (hh.south)--  (fa.north);
   \draw[gray, bend left=90, ultra thick,->, dotted]  (hc.west) to [swap] (hb.west);
\end{tikzpicture}
\caption{The results transfer from the device (the GPU) to the host (CPU).}
\label{fig:Ex3P5}
\end{figure}
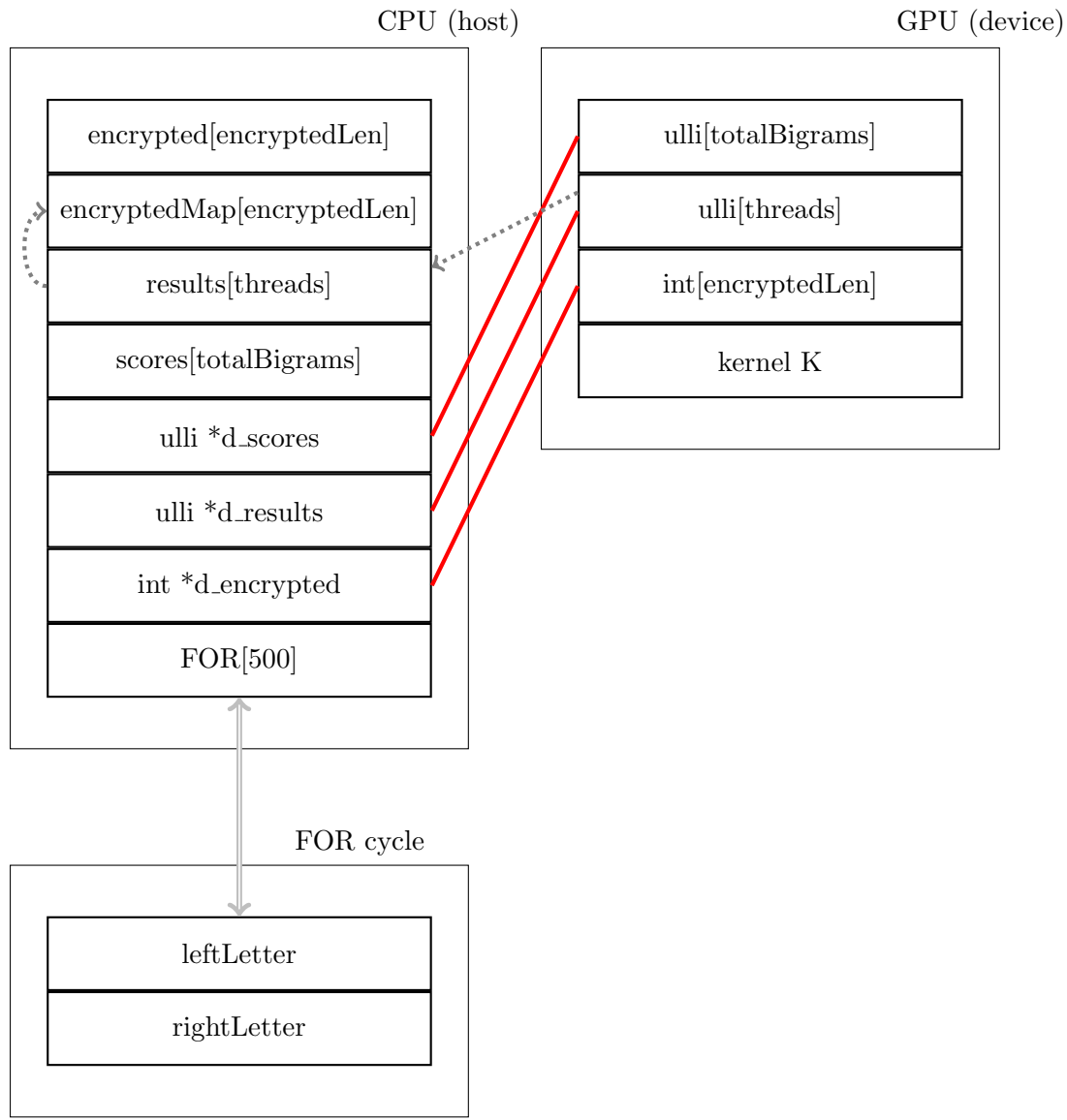

Figure \ref{fig:Ex3P5} summarizes the current snapshot of memory blocks of the host and the device.

We transfer the results from the device (the GPU) to the host (CPU) (line 54 in Listing \ref{list:ex3}), to further analyze them and pick the best candidate. We further update the current text by using the helper function \textbf{climb}. Once this is done, we step into the next iteration of \textbf{for} cycle.

The final construction of the kernel and the complete version of the program is given in Listing \ref{list:complete}. For a better overview some sections are omitted. The full code, including a real example to decrypt, is given in Listing \ref{list:FirstMASCipher}.

\begin{lstlisting}[label={list:complete}, caption=The complete GPU monoalphabetic-substitution solver. , style=customc, numbers=left,stepnumber=1]
#include <stdio.h>
#include <stdlib.h>
#include <string.h>
#include <time.h>

__host__ void extractBigrams() { /* omitted */ }
__host__ int getMaxElement() { /* omitted */ }
__host__ void translateIndexToLetters() { /* omitted */ }
__host__ void climb() { /* omitted */ }
__host__ void demap() { /* omitted */ }

__global__ void K(unsigned long long int *d_scores, int *d_encrypted, int encryptedLen, int leftLetter, int rightLetter, unsigned long long int *d_results){
	int tempLeft = 0;
	int tempRight = tempLeft+1;
	int remainder = threadIdx.x;
	
	while (remainder > 0) {
		tempRight += 1;
		remainder -= 1;
		if (tempRight == 26) {
			tempLeft += 1;
			tempRight = tempLeft + 1;
		}
	}
	
	unsigned long long int total = 0;
	int detectorLeft = 0;
	int detectorRight = 0;
	for (int j=0; j<encryptedLen-1; ++j)
		{
			detectorLeft = d_encrypted[j];
			detectorRight = d_encrypted[j+1];
			if (d_encrypted[j] == leftLetter)
				detectorLeft = tempLeft;
			else if (d_encrypted[j] == tempLeft)
				detectorLeft = leftLetter;
			else if (d_encrypted[j] == rightLetter)
				detectorLeft = tempRight;
			else if (d_encrypted[j] == tempRight)
				detectorLeft = rightLetter;
			else if (d_encrypted[j+1] == leftLetter)
				detectorRight = tempLeft;
			else if (d_encrypted[j+1] == tempLeft)
				detectorRight = leftLetter;
			else if (d_encrypted[j+1] == rightLetter)
				detectorRight = tempRight;
			else if (d_encrypted[j+1] == tempRight)
				detectorRight = rightLetter;
			total += d_scores[detectorLeft*26 + detectorRight];
		}		
	d_results[threadIdx.x] = total;	
	if ((tempLeft == rightLetter) || (tempRight == leftLetter))
		d_results[threadIdx.x] = 0;
}

int main() { /* omitted */ }
\end{lstlisting}

Lines 13-24 (to be found inside the kernel definition) are mapping each thread number to specific unique pair of letters by using the bijective mapping $\Upsilon$. Then, each thread applies the corresponding interchange to calculate the score of the newly created candidate (lines 26-51). Furthermore, we make sure the restriction we have introduced in step 3 of the multi-threaded hill climbing variant is valid (lines 52 and 53). If we happen to be in a thread, where this restriction is not valid, we give the candidate the lowest possible score of 0.

\begin{figure}[H]
\centering
\includegraphics[width=300px]{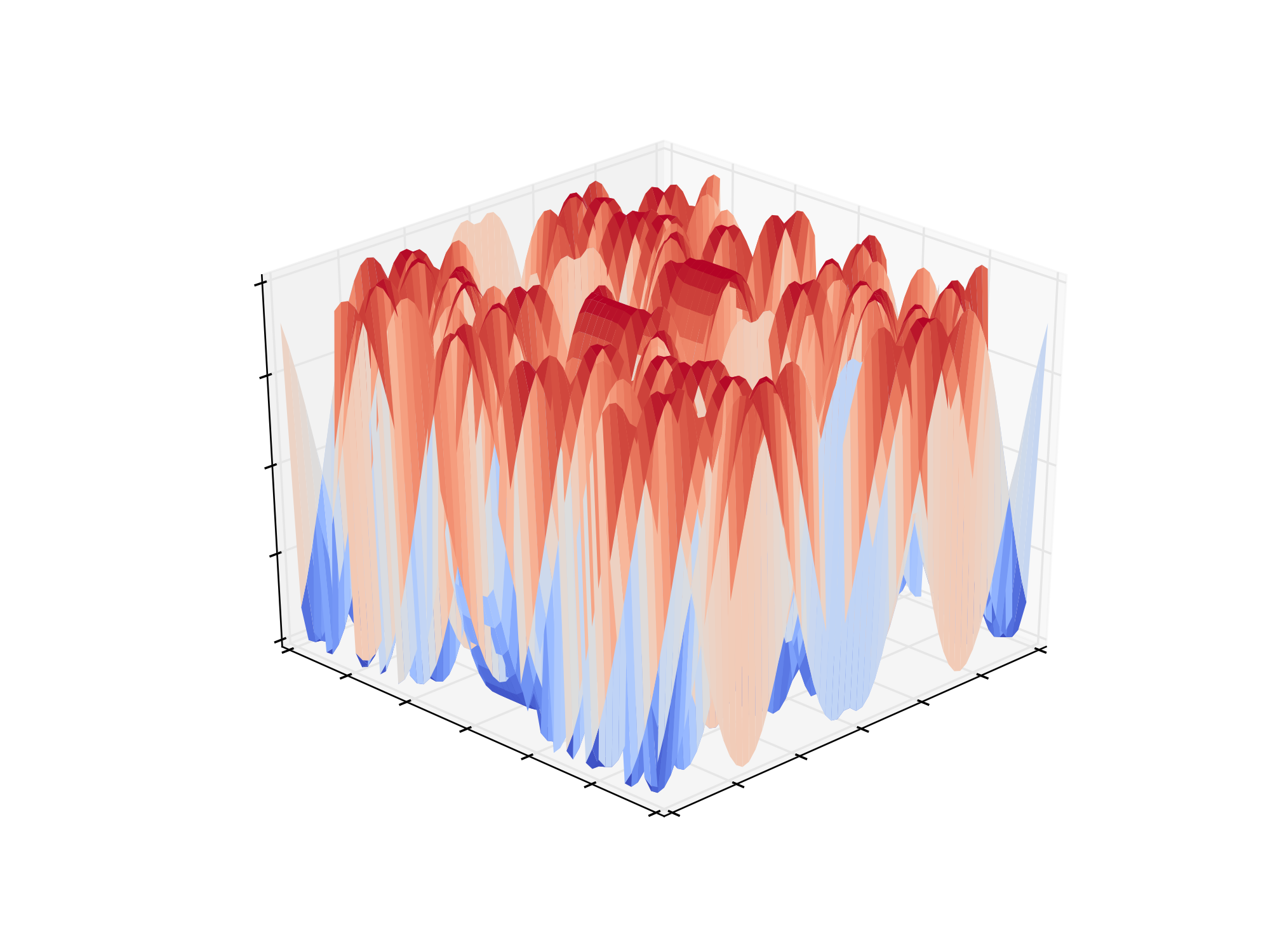} 
\caption{Example of metaheuristics topology. The depicted topology serves as an illustration only. This specific example is generated by using the score (fitness) equation $z = \sin{\left( \vert x \vert + y^8 \right)}$ }
\label{fig:ExampleTopology} 
\end{figure}

It should be emphasized that running a single instance of the final program doesn't guarantee a successful decryption of the encrypted text. In fact, it's highly unlikely that the plaintext is found immediately after the first try. Due to the high number of possible keys in the key search space, i.e. $26!$, the multi-threaded hill-climbing algorithm can easily stuck in some local maximum. A helpful illustration is given in Figure \ref{fig:ExampleTopology}.

The cold areas (nuances of blue) corresponds to candidates with low score values, while the hot areas (nuances of red) corresponds to candidates with higher score values. The algorithm usually starts from some random blue position and
repetitively makes its way to the top. However, the finally reached peak only guarantee that there is no neighbor with a better score value than the peak itself -- it doesn't guarantee that it is the highest peak possible. As you can see in Figure \ref{fig:ExampleTopology} we have a rich set of peaks, but in most real-world application, there is only one peak we are interested in. In literature, each peak is defined as a local maximum, while the highest peak is declared as the global maximum. 

If we are stuck in a local maximum, we should either try to escape from this peak (by actions which worsen the objective function value), or to reinitialize the optimization problem from the beginning. Having this in mind, we should reinitialize the GPU program several times and to collect all the reached peaks. Hopefully, one of them will be the global maximum.

As an exercise, instead of manually reinitialize the program until the desired peak is reached,  implement another loop wrapper of the main loop, to automatically reinitialize the multi-thread hill climbing algorithm. 

Few words should be mentioned regarding some difficulties you could experience during the compilation of a given CUDA source. More precisely, in those cases when you are using Windows OS as a host, the Microsoft provided compilers and linkers do not support the C programming language standard C99 (ISO/IEC 9899:1999). There are some noticeable differences between C99 and newer standards. More detailed information could be found in \cite{cstandards}. 

One common issue which arises during the process of migrating a C99 complaint program  to a newer standard is the restriction on declaring static arrays with non-constant lengths. Having this in mind, all the source codes to be found in this tutorial are attuned to be C99, C11 and C17 compatible. 

One last thing before delving into our first real problem task. As we have already mentioned, all non-CUDA compilation steps are forwarded to a C++ host compiler. If you want to send some addition options to the compiler, like optimization flags, the standard that should be taken under consideration, or anything else, the \textbf{-Xcompiler} switch could be used. More detailed information could be found in the CUDA Toolkit Documentation \cite{cudastandard}.

In Listing \ref{list:FirstMASCipher} the complete source code of Listing \ref{list:complete} is given. This is our first practical usage of the tools we have created so far. The lines 8-11 introduce the flexibility of the provided example, so to be easily compiled on C99, C11 or C17 programming language standards. The source code is to be compiled by \textbf{nvcc}. Can you recover the encrypted message and find the original book from which the plaintext was extracted? Please note: it is highly unlikely that you will be able to successfully decrypt the encrypted message from the first try. A little bit more tries are required for reaching the real solution. In most of the times you will be stuck in some local maximum having a gibberish-looking text. Just keep trying until some reasonable text uncovers.

\pagebreak

\begin{lstlisting}[label={list:FirstMASCipher}, caption=Example of a complete GPU monoalphabetic-substitution solver., style=customc, numbers=left,stepnumber=1, breaklines=true]
#include <stdio.h>
#include <stdlib.h>
#include <string.h>
#include <time.h>
#include <cuda.h>
#include <curand_kernel.h>

#define encrypted "btjpxrmlxpcuvamlxicvjpibtwxvrcimlm trpmtnmtnyvcjxcdxvmwmbtrjjpxamtngxrjbahuqctjpxqgmrjxv cijpxymggcijpxhbtwrqmgmaxmtnjpxhbtwrmyjpxqmvjcijpxpmt njpmjyvcjxjpxtjpxhbtwracutjxtmtaxymrapmtwxnmtnpbrjpcu wpjrjvcufgxnpblrcjpmjjpxscbtjrcipbrgcbtryxvxgccrxnmtn pbrhtxxrrlcjxctxmwmbtrjmtcjpxvjpxhbtwavbxnmgcunjcfvbt wbtjpxmrjvcgcwxvrjpxapmgnxmtrmtnjpxrccjprmexvrmtnjpxh btwrqmhxmtnrmbnjcjpxybrxlxtcifmfegctypcrcxdxvrpmggvxm njpbryvbjbtwmtnrpcylxjpxbtjxvqvxjmjbctjpxvxcirpmggfxa gcjpxn"
#define encryptedLen ((int)sizeof(encrypted)-1)
#define threads 325
#define totalBigrams 676

// a host function to extract and map 
// all the bigrams from a given file
__host__ void extractBigrams(unsigned long long int *scores) {
	FILE* bigramsFile = fopen("bigramParsed", "r");
	while(1){
		char tempBigram[2];
		unsigned long long int tempBigramScore = 0;
		if (fscanf(bigramsFile, "%s %llu", tempBigram, &tempBigramScore) < 2)
			 { break; } 
		scores[(tempBigram[0]-'a')*26 + tempBigram[1]-'a'] = tempBigramScore; 
	}
	fclose(bigramsFile);
}


// a helper function to return the maximum element
__host__ int getMaxElement(unsigned long long int *scores, int length) {
	unsigned long long int tempMaxScore = 0;
	int tempIndex = 0;	
	for (int j=0; j<length; ++j) {
		if (scores[j] > tempMaxScore) {
			tempIndex = j;
			tempMaxScore = scores[j];
		} 
	}	
	return tempIndex;
}


// translate a given index to a map of pair of distinct letters 
__host__ void translateIndexToLetters(int index, int* left, int* right) {

	int tempLeft = 0;
	int tempRight = tempLeft+1;
	
	while (index > 0) {
		tempRight += 1;
		index -= 1;
		if (tempRight == 26) {
			tempLeft += 1;
			tempRight = tempLeft + 1;
		}
	}
	
	*left = tempLeft;
	*right = tempRight;
}


// interchange left with maxLeft and right with maxRight
__host__ void climb(int* encrMap, int length, int left, int maxLeft, int right, int maxRight) {
	for (int j=0; j<length; ++j) {
		if (encrMap[j] == left)
			encrMap[j] = maxLeft;
		else if (encrMap[j] == maxLeft)
			encrMap[j] = left;
		else if (encrMap[j] == right)
				encrMap[j] = maxRight;
		else if (encrMap[j] == maxRight)
				encrMap[j] = right;
	}
}


// demap and print a given array of integer letters
__host__ void demap(int* encrMap, int length) {
	for (int j=0; j<length; ++j) {
		printf("%c", encrMap[j] + 'a');
	}
	printf("\n");
}

__global__ void K(unsigned long long int *d_scores, int *d_encrypted, int leftLetter, int rightLetter, unsigned long long int *d_results){
	int tempLeft = 0;
	int tempRight = tempLeft+1;
	int remainder = threadIdx.x;
	
	while (remainder > 0) {
		tempRight += 1;
		remainder -= 1;
		if (tempRight == 26) {
			tempLeft += 1;
			tempRight = tempLeft + 1;
		}
	}
	
	unsigned long long int total = 0;
	int detectorLeft = 0;
	int detectorRight = 0;
	for (int j=0; j<encryptedLen-1; ++j)
		{
			detectorLeft = d_encrypted[j];
			detectorRight = d_encrypted[j+1];
			if (d_encrypted[j] == leftLetter)
				detectorLeft = tempLeft;
			else if (d_encrypted[j] == tempLeft)
				detectorLeft = leftLetter;
			else if (d_encrypted[j] == rightLetter)
				detectorLeft = tempRight;
			else if (d_encrypted[j] == tempRight)
				detectorLeft = rightLetter;
			else if (d_encrypted[j+1] == leftLetter)
				detectorRight = tempLeft;
			else if (d_encrypted[j+1] == tempLeft)
				detectorRight = leftLetter;
			else if (d_encrypted[j+1] == rightLetter)
				detectorRight = tempRight;
			else if (d_encrypted[j+1] == tempRight)
				detectorRight = rightLetter;
			total += d_scores[detectorLeft*26 + detectorRight];
		}		
	d_results[threadIdx.x] = total;	
	if ((tempLeft == rightLetter) || (tempRight == leftLetter))
		d_results[threadIdx.x] = 0;
}

int main() 
{ 

	srand(time(0));
	int encryptedMap[encryptedLen];
	
	unsigned long long int results[threads];
	
	// we map the encrypted message to integers
	for (int j=0; j<encryptedLen; ++j)
		encryptedMap[j] = encrypted[j] - 'a';
	
	unsigned long long int scores[totalBigrams];	
	extractBigrams(scores);
		
	int size = sizeof(unsigned long long int);
	unsigned long long int *d_scores;
	unsigned long long int *d_results;
	int *d_encrypted;
	
	// allocate memory blocks on the device
	cudaMalloc((void **)&d_scores, size*totalBigrams);
	cudaMalloc((void **)&d_results, size*threads);
	cudaMalloc((void **)&d_encrypted, sizeof(int)*encryptedLen);	
	
	// copy the scores array to the device memory (only once)
	cudaMemcpy(d_scores, scores, size*totalBigrams, cudaMemcpyHostToDevice);
	
	int leftLetter = 0;
	int rightLetter = 0;
	int tempMaxIndex = 0;
	int tempMaxLeft = 0;
	int tempMaxRight = 0;
	unsigned long long int maxScore = 0;
	
	for (int j=0; j<500; ++j) 	{
	
		// pick two random distinct letters; we make sure they do exists in the encrypted text
		leftLetter = encrypted[rand() % encryptedLen] - 'a';
		rightLetter = leftLetter;
		while (rightLetter == leftLetter)
			rightLetter = encrypted[rand() % encryptedLen] - 'a';
	
		// copy the encrypted msg array to the device memory (after each climbing)
		cudaMemcpy(d_encrypted, encryptedMap, sizeof(int)*encryptedLen, cudaMemcpyHostToDevice);
	
	
		K<<<1,threads>>>(d_scores, d_encrypted, leftLetter, rightLetter, d_results);
		cudaDeviceSynchronize();
	
		// copy the results array back to the host memory
		cudaMemcpy(results, d_results, size*threads, cudaMemcpyDeviceToHost);
	
		tempMaxIndex = getMaxElement(results, threads);
		
		if (results[tempMaxIndex] > maxScore) {
			maxScore = results[tempMaxIndex];
			tempMaxLeft = 0;
			tempMaxRight = 0;
			translateIndexToLetters(tempMaxIndex, &tempMaxLeft, &tempMaxRight);
			climb(encryptedMap, encryptedLen, leftLetter, tempMaxLeft, rightLetter, tempMaxRight);
			printf("%d %llu\n", tempMaxIndex, results[tempMaxIndex]);
			printf("%d %d %d %d\n", tempMaxLeft, tempMaxRight, leftLetter, rightLetter);
			demap(encryptedMap, encryptedLen);
			
		}
	}
	
	// we free the allocated device memory 
	cudaFree(d_scores);
	cudaFree(d_encrypted);
	cudaFree(d_results);
	
	return 0;	
}
\end{lstlisting}

\section{CUDA blocks and pseudo-randomness}
\label{sec:CUDABlocks} 

So far we have utilized the thread parameter inside a given kernel launch only. However, as we will later see, this is unwise approach due to inefficient usage of resources. Furthermore, the CUDA platform has a limitation of the maximum threads we can initiate inside a CUDA block -- 512, 1024 or 2048. Those limits depend from the specific model of the device.

\subsection{CUDA blocks}

\begin{figure}[h]
\centering
\includegraphics[width=250px]{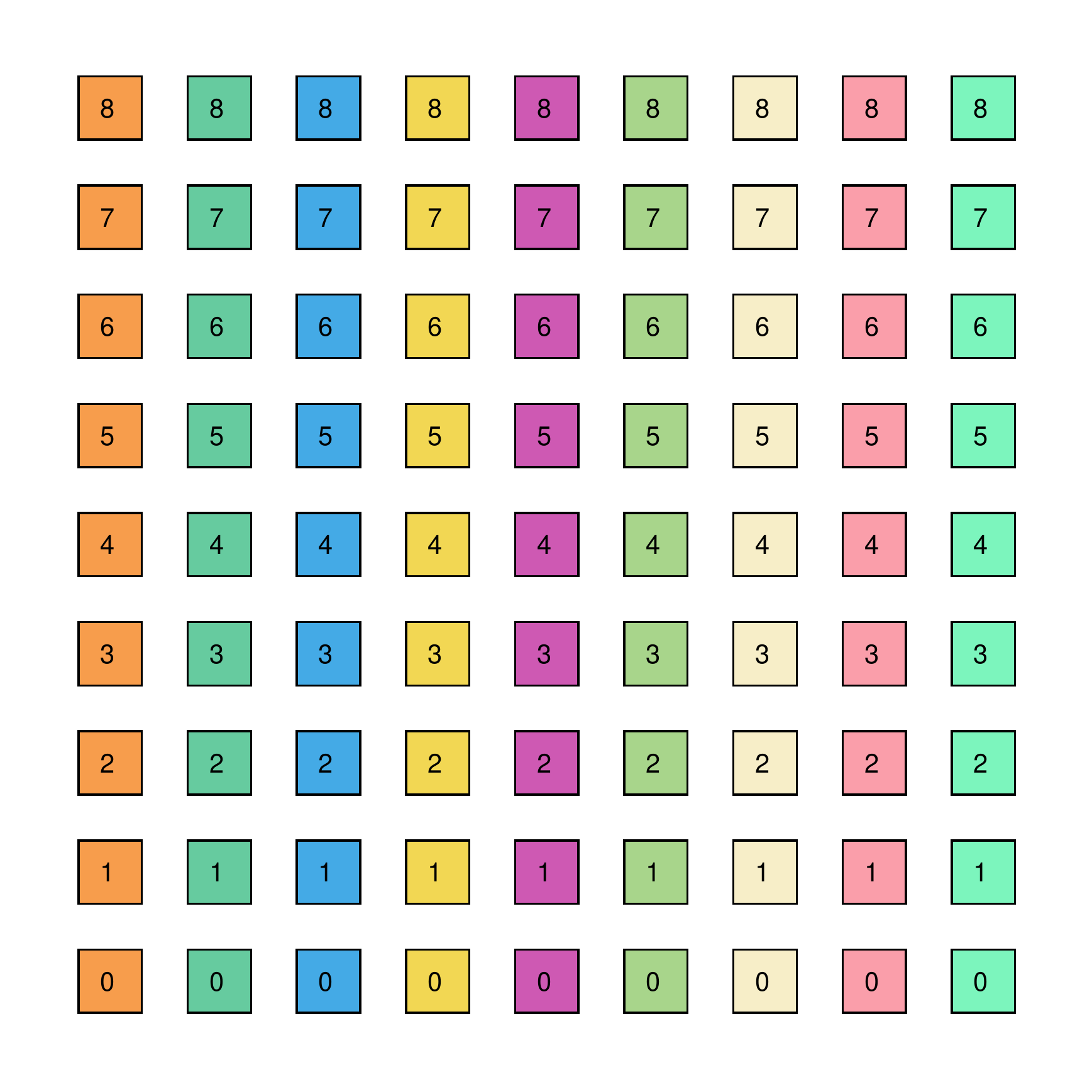} 
\caption{Example of GPU kernel organized by blocks}
\label{fig:ExampleBlocks} 
\end{figure}

A CUDA block, in short, is a collection of threads. For example, we can have $x$ blocks with $y$ threads each or a total of $xy$ threads. A visual example for a CUDA kernel organized by blocks is given in Figure \ref{fig:ExampleBlocks}. Each column is depicted with a different color since it corresponds to a distinct kernel block. However, each thread number is unique only inside a given block. Having this in mind, we should use some other CUDA feature.

 To distinguish, for example, the following two threads: the first one with number $i$ coming from the block $b_1$; and the second one with the same number $i$ initiated from another block $b_2$. For this, we are going to use the built-in variables \textbf{blockDim} (returns the size of a block) and \textbf{blockIdx} (returns the unique identification number of the current block).

Let's say we want to launch a kernel having 16 blocks with 8 threads each. The task that should be performed involves a simple power calculation. Assume that we would like to raise the numbers from 0 to 7 (the thread IDs) to the powers 0 to 15 (the block IDs). To achive this, we can use the following code (see Listing \ref{list:blocks}).

\begin{minipage}{\linewidth}
\begin{lstlisting}[label={list:blocks}, caption=Example of a function organized by blocks, style=customc, numbers=left,stepnumber=1]
#include <stdio.h>

__device__ int powr(int x, int n) {
	int res = x;
	for (int i=n; i>1; --i)
		res *= x;
	return (n==0)?1:res;
}

__global__ void L(){
	if (threadIdx.x != 0 || blockIdx.x != 0)
		printf("%d powered to %d is: %d \n", threadIdx.x, blockIdx.x, powr(threadIdx.x, blockIdx.x));
}

int main() {
	L<<<8,16>>>();
	cudaDeviceSynchronize();
	return 0;
}
\end{lstlisting}
\end{minipage}

Lines 3-8 define a helper function \textbf{powr} (x,n), which raises $x$ to the power of $n$, where $x$ and $n$ are integers. The kernel function is specified at the lines 10-13. The actual calling of the kernel is made at line 16. As you can see we launch the kernel with parameters \textbf{8,16}, i.e. 8 blocks with 16 threads each. 

Now, we can divide or parallelize a given problem to blocks and threads. However, if you recall the \ac{mas} problem (Section \ref{sec:GPUSubstAnalysis}), we have used the \ac{prng} of the host only. This leaded to a noticeable overhead and over-complication of the problem: we first generated two pseudo-random letters $\psi_i$ and $\psi_j$, then we pass them to the device as kernel parameters, to further interchange them with all the possible unordered pairs of letters in the corresponding alphabet.

\subsection{CUDA as pseudo-random number generator}
\label{subsec:pseudoRandom}

What if we want to use a \ac{prng} inside the device? Furthermore, how to be sure that the pseudo-randomly generated numbers inside the threads are not equal? For this a fast and reliable \ac{prng} is needed. Using the host \ac{prng} routines is not a wise approach -- we have to repeatedly transfer the generated values back and forth from the host to the device.

With the help of the following CUDA code example given in Listing \ref{list:pseudoDevice}, we are going to learn some new techniques:
\begin{itemize}
\item{How to define and launch multiple kernels inside a single CUDA program.}
\item{How to use the external libraries provided by CUDA.}
\item{How to generate random float numbers by using the device itself.}
\end{itemize}

\begin{lstlisting}[label={list:pseudoDevice}, caption=Parallel CUDA pseudo-random float number generator, style=customc, numbers=left,stepnumber=1]
#include <stdio.h>
#include <cuda.h>
#include <curand_kernel.h>

#define B 8
#define T 8

__global__ void setupKernel(curandState* state, unsigned long seed) {
	int idx = blockIdx.x*blockDim.x + threadIdx.x;
	curand_init(seed, idx, 0, &state[idx]);
}

__global__ void generate(curandState* globalState, float* devRandomValues) {
	int idx = blockIdx.x*blockDim.x + threadIdx.x;
	curandState localState = globalState[idx];
	float randF = curand_uniform(&localState);
	devRandomValues[idx] = randF;
	globalState[idx] = localState;
}

int main() {
	curandState* devStates;
	float* randomValues = new float[B*T];
	float* devRandomValues;
	
	// we allocate the required memory blocks
	cudaMalloc(&devStates, B*T*sizeof(curandState));
	cudaMalloc(&devRandomValues, B*T*sizeof(*randomValues));
	
	// initial setup
	setupKernel<<<B,T>>>(devStates, time(NULL));	
	cudaDeviceSynchronize();
	
	// generator
	generate<<<B,T>>>(devStates, devRandomValues);	
	cudaDeviceSynchronize();
	
	// extract the random values from the device
	cudaMemcpy(randomValues, devRandomValues, B*T*sizeof(*randomValues), cudaMemcpyDeviceToHost);
	
	// feedback
	for (int i=0;i<B*T;++i)
		printf("%f\n", randomValues[i]);
	
	// cleaning up
	cudaFree(devRandomValues);
	cudaFree(devStates);
	delete randomValues;
	
	return 0;
}
\end{lstlisting}

First (lines 2-3), we include the CUDA library, as well as the CUDA random kernel headers. For more flexibility, we define the block and thread numbers (lines 5-6) as constants \textbf{B} and \textbf{T}. 

The first kernel, which can be found at lines 8-11, is initializing each thread (from the pool of threads) with a unique initial seed. To achieve this we are using the built-in CUDA function \textbf{curand\_init}. We save each such seed to the corresponding cell inside the array \textbf{state}. 

The second kernel, which is defined at lines 13-19, uses the seed values initialized by the first kernel and the CUDA built-in function \textbf{curand\_uniform} to generate a random float number in the interval $\left[ 0.0, 1.0 \right]$. We save this value in a corresponding cell inside an array named \textbf{devRandomValues}. Furthermore, we update the seed corresponding to this thread, so each consequent \ac{prng} yields a different result.

The main function follows the logic introduced during the examples in the previous sections. At lines 27-28 we allocate the device memory blocks:
\begin{itemize}
\item{a pointer to a device memory block \textbf{devStates}, which is going to save our current thread seeds;}
\item{a device memory block \textbf{devRandomValues}, which is going to save the generated pseudo-random values;}
\end{itemize}

Then, we launch the first kernel (line 31) to initialize the seeds, which is followed by the launch of the second kernel (line 35) to populate the pseudo-randomly generated numbers inside the memory block \textbf{devRandomValues}. The remaining lines were already discussed in the previous examples.

Now, we need to address the following questions:
\begin{itemize}
\item{How to generate pseudo-random integer numbers in a predefined interval?}
\item{How to utilize the parallel device \ac{prng}s for automatic cryptanalysis?}
\end{itemize}

\subsection{A stand-alone CUDA application to automatically attack MAS ciphertexts}
\label{sub:standalone}

We have all the necessary instruments to construct a complete stand-alone CUDA \ac{mas} cipher cryptanalysis tool. We will proceed as follows:

\begin{enumerate}
\item{Setup the CUDA device and initialize a seed for each thread of the GPU.}
\item{Make a local copy of the encrypted text inside each thread.}
\item{Initiate a simple hill climbing routine inside each thread by using bigrams. The \ac{prng}s are launched inside the threads. All \ac{prng}s are launched independently to guarantee a feed of unique values to the corresponding thread.}
\item{After passing some predefined threshold value to the hill climbing algorithm, the thread will report back the reached peak value. Then, the results are collected and reported back to the host.} 
\item{The host compares the final scores of all candidates and announces the best one.}
\end{enumerate}

Listing \ref{list:multiHill} contains the complete CUDA source code of the program.


\begin{lstlisting}[label={list:multiHill}, caption=Parallel CUDA monoalphabetic solver with pseudo-random float number generator nested inside the device, style=customc, numbers=left,stepnumber=1]
#include <stdio.h>
#include <stdlib.h>
#include <string.h>
#include <cuda.h>
#include <curand_kernel.h>

#define encrypted "lwzlmomwmogfeohitkldabztdartw hgylwzlmomwmtrstmmtklfwdtkaslegfxtfmogfaslouflgk egdzofamogflgyassmikttafrywkmitkdgktygkalofustst mmtkgymitgkouofasmtvmmitktdabztlwzlmomwmtralofus tstmmtkfwdtkasgkloufgkmcggkdgktgytaeigkacigstcgk rgkukgwhgyyouwktlegdzofamogfgyegfxtfmogfaslouflg kegdzofamogflgyassmikttgymitlttstdtfml"
#define encryptedLen ((int)sizeof(encrypted)-1)

#define B 16
#define T 16
#define CLIMBINGS 5000
#define ALPHABET 26
#define totalBigrams ((int)ALPHABET*ALPHABET)

// a host function to extract and map all the bigrams from a given file
__host__ void extractBigrams(unsigned long long int *scores) {
	FILE* bigramsFile = fopen("bigramParsed", "r");
	while(1){
		char tempBigram[2];
		unsigned long long int tempBigramScore = 0;
		if (fscanf(bigramsFile, "%s %llu", tempBigram, &tempBigramScore) < 2)
			 { break; } 
		scores[(tempBigram[0]-'a')*ALPHABET + tempBigram[1]-'a'] = tempBigramScore; 
	}
	fclose(bigramsFile);
}

// demap and print a given array of integer letters
__host__ __device__ void demap(int* encrMap, int length) {
	for (int j=0; j<length; ++j) {
		printf("%c", encrMap[j] + 'a');
	}
	printf("\n");
}

// calculate a bigram score of a given decrypted (mapped) candidate 
__host__ unsigned long long int candidateScore(int* decrMsg, int length, unsigned long long int* scores) {
	unsigned long long int total = 0;
	for (int j=0; j<length-1; ++j) {
		total += scores[ALPHABET*decrMsg[j] + decrMsg[j+1]];
	}
	return total;
}

__global__ void setupKernel(curandState* state, unsigned long seed) {
	int idx = blockIdx.x*blockDim.x + threadIdx.x;
	curand_init(seed, idx, 0, &state[idx]);
}

__global__ void K(unsigned long long int *d_scores, int *d_encrypted, curandState* globalState, int *d_decrypted){
	int* localBufferEncrypted = new int[encryptedLen];
	for (int j=0; j<encryptedLen; ++j)
		localBufferEncrypted[j] = d_encrypted[j];
	int idx = blockIdx.x*blockDim.x + threadIdx.x;
	curandState localState = globalState[idx];

	long long int delta = 0;
	float randF = 0;
	int leftLetter = 0;
	int rightLetter = 0;
	int pair[2];
	
	for (int cycles=0; cycles<CLIMBINGS; ++cycles) {	
		delta = 0;		
		randF = curand_uniform(&localState);
		leftLetter = randF*ALPHABET; 	
		rightLetter = leftLetter;
		while (rightLetter == leftLetter) {
			randF = curand_uniform(&localState);
			rightLetter = randF*ALPHABET; 
		}
		
		for (int j=0; j<encryptedLen-1; ++j)
			{
				pair[0] = localBufferEncrypted[j];
				pair[1] = localBufferEncrypted[j+1];
				delta -= d_scores[ALPHABET*pair[0] + pair[1]];
				
				if (pair[0] == leftLetter)
					pair[0] = rightLetter;
				else if (pair[0] == rightLetter)
					pair[0] = leftLetter;
					
				if (pair[1] == leftLetter)
					pair[1] = rightLetter;
				else if (pair[1] == rightLetter)
					pair[1] = leftLetter;					
				delta += d_scores[ALPHABET*pair[0] + pair[1]];				
			}			
			
			if (delta > 0) {
				for (int j=0; j<encryptedLen; ++j) {
					if (localBufferEncrypted[j] == leftLetter)
						localBufferEncrypted[j] = rightLetter;
					else if (localBufferEncrypted[j] == rightLetter)
						localBufferEncrypted[j] = leftLetter;
				}
			}
		}
	
	for (int j=0; j<encryptedLen; ++j)
		d_decrypted[idx*encryptedLen+j] = localBufferEncrypted[j];
	delete localBufferEncrypted;
}

int main() {
	
	int threads = B*T;
	int encryptedMap[encryptedLen];
	
	// we map the encrypted message to integers
	for (int j=0; j<encryptedLen; ++j)
		encryptedMap[j] = encrypted[j] - 'a';
	
	unsigned long long int scores[totalBigrams];	
	extractBigrams(scores);
		
	unsigned long long int *d_scores;
	int *d_encrypted, *d_decrypted;
	int* decrypted = new int[encryptedLen*threads];
	
	// allocate memory blocks on the device
	cudaMalloc((void **)&d_scores, sizeof(unsigned long long int)*totalBigrams);
	cudaMalloc((void **)&d_encrypted, sizeof(int)*encryptedLen);
	cudaMalloc((void **)&d_decrypted, sizeof(int)*encryptedLen*threads);	
	
	// copy the scores array and encrypted msg array to the device memory
	cudaMemcpy(d_scores, scores, sizeof(unsigned long long int)*totalBigrams, cudaMemcpyHostToDevice);
	cudaMemcpy(d_encrypted, encryptedMap, sizeof(int)*encryptedLen, cudaMemcpyHostToDevice);
	
	// we allocate the required memory blocks for pseudo-random generation routines
	curandState* devStates;
	cudaMalloc(&devStates, B*T*sizeof(curandState));
	
	// initial setup
	setupKernel<<<B,T>>>(devStates, time(NULL));	
	cudaDeviceSynchronize();

	
    // the "flock" of threads initiating automatic cryptanalysis in parallel
	K<<<B,T>>>(d_scores, d_encrypted, devStates, d_decrypted);
	cudaDeviceSynchronize();

	// copy the results array back to the host memory
	cudaMemcpy(decrypted, d_decrypted, sizeof(int)*encryptedLen*threads, cudaMemcpyDeviceToHost);
	
	int bestCandidate = 0;
	unsigned long long int bestScore = 0;
	
	// we extract the best candidate among all threads
	for (int j=0; j<threads; ++j)	{
		unsigned long long int currentScore = candidateScore(&decrypted[encryptedLen*j], encryptedLen, scores);
		if (currentScore > bestScore) {
			bestScore = currentScore;
			bestCandidate = j;
		}		
	}	
	
	// final solution output
	printf("Best candidate score: %llu\n", bestScore);
	demap(&decrypted[encryptedLen*bestCandidate], encryptedLen);		
			
	// we free the allocated device and host memory blocks
	cudaFree(d_scores);
	cudaFree(d_encrypted);
	cudaFree(d_decrypted);
	cudaFree(devStates);
	delete decrypted;
	
	return 0;
	
}
\end{lstlisting}

Now, let's trace the program behavior by employing the useful memory snapshot approach. At lines 7-14 we define several constants we are going to use throughout the program logic. A summary of all the constants is given in Table \ref{tab:magicConstants}.

\begin{table}[h]
  \begin{center}
  	\caption{Overview of the predefined constants}
  	\label{tab:magicConstants}
    \begin{tabularx}{\linewidth}{@{} r|X @{}}      
      \hline
      \textbf{Constant} & \textbf{Logic} \\
      \hline
      \hline
      encrypted & the encrypted text \\
      encryptedLen & the length of the encrypted text \\
      B & the number of blocks \\
      T & the number of threads \\
      CLIMBINGS & the threshold value of hill climbing tries for each distinct thread \\
      ALPHABET & the total count of letters in the current alphabet \\
      totalBigrams & the total count of distinct bigrams \\
	  \hline	
    \end{tabularx}
    \end{center}	
\end{table}

\begin{figure}[H]
\centering
\begin{tikzpicture}
  \node[block3] (ha) {encrypted[LL]};
  \node[block3, below=0cm of ha] (hb) {encryptedMap[LL]};
  \node[block3, below=0cm of hb] (hc) {scores[totalBigrams]};
  \node[block3, below=0cm of hc] (hd) {ulli *d\_scores};
  \node[block3, below=0cm of hd] (he) {int *d\_encrypted};
  \node[block3, below=0cm of he] (hf) {int *d\_decrypted};
  \node[block3, below=0cm of hf] (hg) {int decrypted [LL*B*T]};
  \node[block3,right=2cm of ha] (da) {empty};
  \node[draw,inner xsep=5mm,inner ysep=7mm,fit=(ha) (hb) (hc) (hd) (he) (hf) (hg) ,label={70:CPU (host)}](f){};
  \node[draw,inner xsep=5mm,inner ysep=7mm,fit=(da) ,label={60:GPU (device)}]{};
\end{tikzpicture}
\caption{Initial pointers and diagram.}
\label{fig:Ex4P1}
\end{figure}
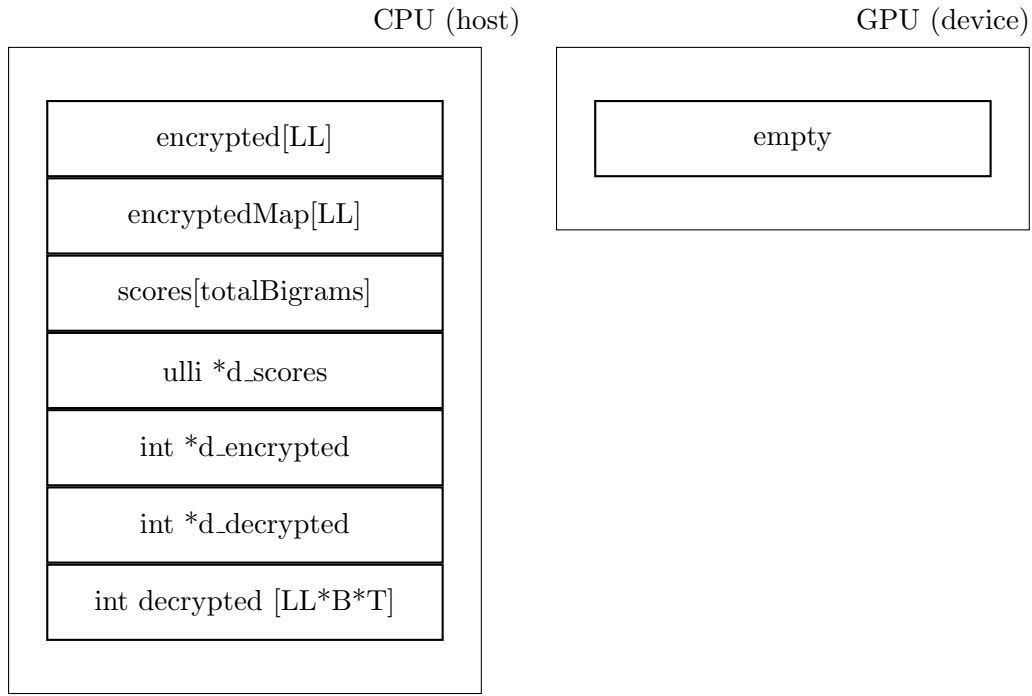

The current snapshot of the host and device memory blocks, up to line 123, is given in Figure \ref{fig:Ex4P1}.

As usual, we setup the necessary pointers. For simplicity, we  denote the variable type \textbf{unsigned long long int} with \textbf{ulli}. Furthermore, we denote the variable \textbf{encryptedLen} as \textbf{LL}.

The functions \textbf{extractBigrams()} (lines 17-27) and \textbf{demap()}  (lines 29-35) are the same as those inside the previous implementation of the \ac{mas} cryptanalysis tool (see Listing \ref{list:FirstMASCipher}). However, we introduce a new function named \textbf{candidateScore} (lines 38-49), which is going to return the bigram score of a given string.

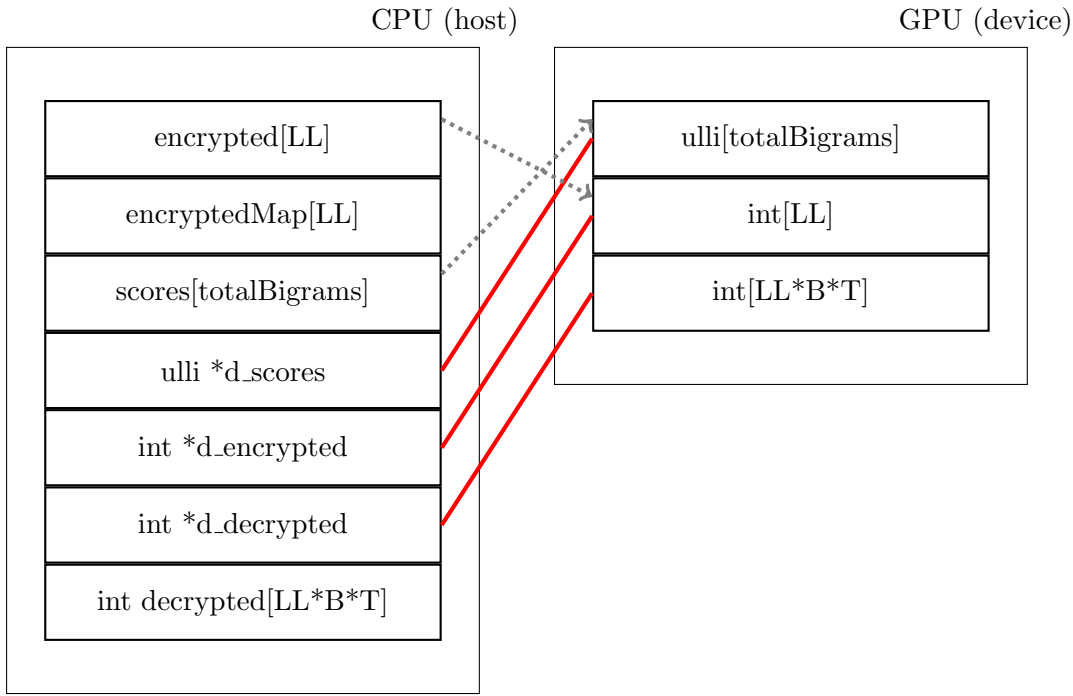
\begin{figure}[H]
\centering
\begin{tikzpicture}
  \node[block3] (ha) {encrypted[LL]};
  \node[block3, below=0cm of ha] (hb) {encryptedMap[LL]};
  \node[block3, below=0cm of hb] (hc) {scores[totalBigrams]};
  \node[block3, below=0cm of hc] (hd) {ulli *d\_scores};
  \node[block3, below=0cm of hd] (he) {int *d\_encrypted};
  \node[block3, below=0cm of he] (hf) {int *d\_decrypted};
  \node[block3, below=0cm of hf] (hg) {int decrypted[LL*B*T]};
  \node[block3,right=2cm of ha] (da) {ulli[totalBigrams]};
  \node[block3,below= 0cm of da] (db) {int[LL]};
  \node[block3,below= 0cm of db] (dc) {int[LL*B*T]};
  \node[draw,inner xsep=5mm,inner ysep=7mm,fit=(ha) (hb) (hc) (hd) (he) (hf) (hg) ,label={70:CPU (host)}](f){};
  \node[draw,inner xsep=5mm,inner ysep=7mm,fit=(da) (db) (dc),label={60:GPU (device)}]{};
   \draw[red, ultra thick, -] (hd.east) -- (da.west);
   \draw[red, ultra thick, -] (he.east)-- (db.west);
   \draw[red, ultra thick, -] (hf.east)-- (dc.west);
   \draw[gray, ultra thick, ->, dotted] ($(hc.south east)!0.75!(hc.north east)$) --  ($(da.south west)!0.75!(da.north west)$) ;
   \draw[gray, ultra thick, ->, dotted] ($(ha.south east)!0.75!(ha.north east)$) --  ($(db.south west)!0.75!(db.north west)$) ;
\end{tikzpicture}
\caption{Allocations and transfers.}
\label{fig:Ex4P2}
\end{figure}

After that, as defined at lines 124-131 we first allocate and then transfer the data needed to be accessible through the GPU kernel routines (see Figure \ref{fig:Ex4P2}).

We first allocate an ulli array, which is going to keep the scores inside the device memory as well. Then, we proceed with allocating two additional int arrays: the first will keep the encrypted text (the second row-block from the device memory block), while the second will keep all the decrypted candidates yielded by all the B*T threads (the third row-block in the device memory). Then, depicted by gray dotted arrows, we transfer the bigrams score from the host to the device , as well as the encrypted message itself.

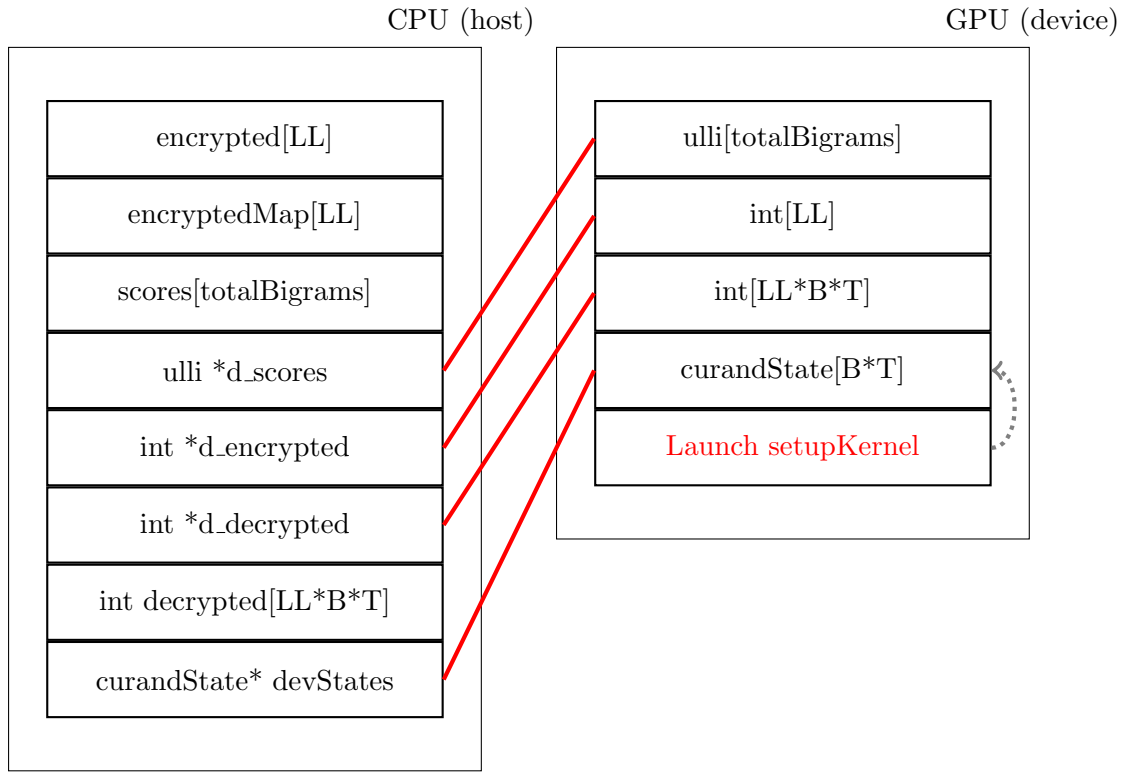
\begin{figure}[H]
\centering
\begin{tikzpicture}
  \node[block3] (ha) {encrypted[LL]};
  \node[block3, below=0cm of ha] (hb) {encryptedMap[LL]};
  \node[block3, below=0cm of hb] (hc) {scores[totalBigrams]};
  \node[block3, below=0cm of hc] (hd) {ulli *d\_scores};
  \node[block3, below=0cm of hd] (he) {int *d\_encrypted};
  \node[block3, below=0cm of he] (hf) {int *d\_decrypted};
  \node[block3, below=0cm of hf] (hg) {int decrypted[LL*B*T]};
  \node[block3, below=0cm of hg] (hh) {curandState* devStates};
  \node[block3,right=2cm of ha] (da) {ulli[totalBigrams]};
  \node[block3,below= 0cm of da] (db) {int[LL]};
  \node[block3,below= 0cm of db] (dc) {int[LL*B*T]};
  \node[block3,below= 0cm of dc] (dd) {curandState[B*T]};
   \node[block3,below= 0cm of dd] (de) {\color{red}{Launch setupKernel}};
  \node[draw,inner xsep=5mm,inner ysep=7mm,fit=(ha) (hb) (hc) (hd) (he) (hf) (hg) (hh) ,label={70:CPU (host)}](f){};
  \node[draw,inner xsep=5mm,inner ysep=7mm,fit=(da) (db) (dc) (dd) (de),label={60:GPU (device)}]{};
   \draw[red, ultra thick, -] (hd.east) -- (da.west);
   \draw[red, ultra thick, -] (he.east)-- (db.west);
   \draw[red, ultra thick, -] (hf.east)-- (dc.west);
   \draw[red, ultra thick, -] (hh.east)-- (dd.west);   
    \draw[gray, bend right=90, ultra thick,->, dotted]  (de.east) to [swap] (dd.east);
\end{tikzpicture}

\caption{Launching the first kernel.}
\label{fig:Ex4P3}
\end{figure}

We are ready to setup the device built-in \ac{prng} (lines 133-138). The current memory block snapshot of the host and the device is given in Figure \ref{fig:Ex4P3}.

We allocate a \textbf{curandState} array, having size B*T equal to the number of device threads, which is going to hold down the initial seeds of the device built-in \ac{prng}. Then, we launch the kernel \textbf{setupKernel} which populates the curandState device array with the provided by the kernel seeds. When a given thread with global index $I$ needs a pseudo-randomly generated number, it directly communicates with cell indexed $I$ inside the curandState device array. Moreover, after each call, the cell with index $I$ is updated to guarantee an infinite stream of pseudo-randomly generated numbers. 

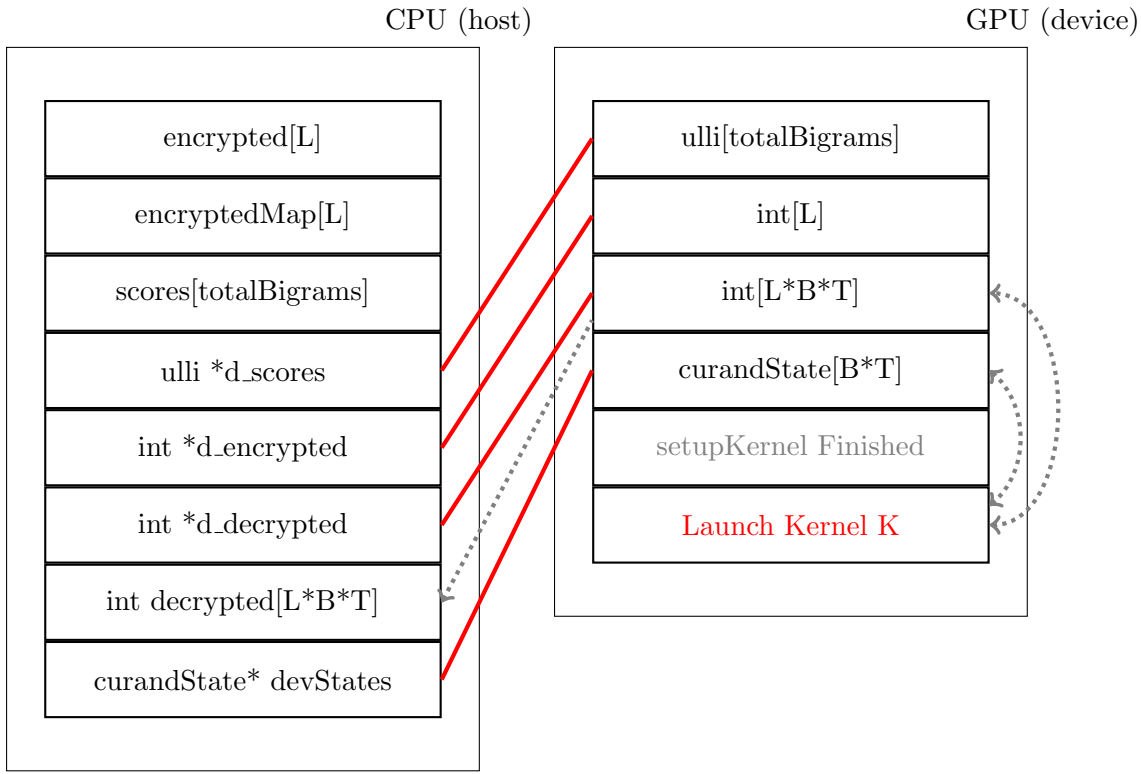
\begin{figure}[H]
\centering
\begin{tikzpicture}
  \node[block3] (ha) {encrypted[L]};
  \node[block3, below=0cm of ha] (hb) {encryptedMap[L]};
  \node[block3, below=0cm of hb] (hc) {scores[totalBigrams]};
  \node[block3, below=0cm of hc] (hd) {ulli *d\_scores};
  \node[block3, below=0cm of hd] (he) {int *d\_encrypted};
  \node[block3, below=0cm of he] (hf) {int *d\_decrypted};
  \node[block3, below=0cm of hf] (hg) {int decrypted[L*B*T]};
  \node[block3, below=0cm of hg] (hh) {curandState* devStates};
  \node[block3,right=2cm of ha] (da) {ulli[totalBigrams]};
  \node[block3,below= 0cm of da] (db) {int[L]};
  \node[block3,below= 0cm of db] (dc) {int[L*B*T]};
  \node[block3,below= 0cm of dc] (dd) {curandState[B*T]};
   \node[block3,below= 0cm of dd] (de) {\color{gray}{setupKernel Finished}};
  \node[block3,below= 0cm of de] (df) {\color{red}{Launch Kernel K}};
  \node[draw,inner xsep=5mm,inner ysep=7mm,fit=(ha) (hb) (hc) (hd) (he) (hf) (hg) (hh) ,label={70:CPU (host)}](f){};
  \node[draw,inner xsep=5mm,inner ysep=7mm,fit=(da) (db) (dc) (dd) (de) (df),label={60:GPU (device)}]{};
   \draw[red, ultra thick, -] (hd.east) -- (da.west);
   \draw[red, ultra thick, -] (he.east)-- (db.west);
   \draw[red, ultra thick, -] (hf.east)-- (dc.west);
   \draw[red, ultra thick, -] (hh.east)-- (dd.west);   
   \draw[gray, bend right=50, ultra thick, <->, dotted]  ($(df.south east)!0.75!(df.north east)$) to [swap] (dd.east);
    \draw[gray, bend right=90, ultra thick, <->, dotted]  (df.east) to [swap] (dc.east);
    \draw[gray, ultra thick, <-, dotted] ($(hg.south east)!0.50!(hg.north east)$) --  ($(dc.south west)!0.15!(dc.north west)$) ;
\end{tikzpicture}
\caption{Launching the second kernel.}
\label{fig:Ex4P4}
\end{figure}

We are ready to launch the hill climbing routine device kernel K (line 142). Next, we transfer back (from the device to the host memory) the local optimums yielded by each device thread. The snapshot is given in Figure \ref{fig:Ex4P4}.

Once the kernel \textbf{setupKernel} finished, we launch the kernel \textbf{K}. Each device thread is working with its own copy of the encrypted text and therefore each thread yields its own local optimum of the decrypted candidate. During the hill climbing routine, each thread is independently generating and utilizing pseudo-random numbers by querying its corresponding curandState cell. Once the thread finished, the respective local optimum is written down in the device memory block (the third row). Finally, all the local optimums are transferred back to the host, by using the dedicated integer array \textbf{decrypted}. 

\begin{figure}[H]
\centering
\begin{tikzpicture}
  \node[block3] (ha) {encrypted[L]};
  \node[block3, below=0cm of ha] (hb) {encryptedMap[L]};
  \node[block3, below=0cm of hb] (hc) {scores[totalBigrams]};
  \node[block3, below=0cm of hc] (hd) {ulli *d\_scores};
  \node[block3, below=0cm of hd] (he) {int *d\_encrypted};
  \node[block3, below=0cm of he] (hf) {int *d\_decrypted};
  \node[block3, below=0cm of hf] (hg) {empty};
  \node[block3, below=0cm of hg] (hh) {curandState* devStates};
  \node[block3,right=2cm of ha] (da) {empty};
  \node[block3,below= 0cm of da] (db) {empty};
  \node[block3,below= 0cm of db] (dc) {empty};
  \node[block3,below= 0cm of dc] (dd) {empty};
   \node[block3,below= 0cm of dd] (de) {\color{gray}{setupKernel Finished}};
  \node[block3,below= 0cm of de] (df) {\color{gray}{K Finished}};
  \node[draw,inner xsep=5mm,inner ysep=7mm,fit=(ha) (hb) (hc) (hd) (he) (hf) (hg) (hh) ,label={70:CPU (host)}](f){};
  \node[draw,inner xsep=5mm,inner ysep=7mm,fit=(da) (db) (dc) (dd) (de) (df),label={60:GPU (device)}]{};
   \draw[red,  thick, -, dotted] (hd.east) -- (da.west);
   \draw[red,  thick, -, dotted] (he.east)-- (db.west);
   \draw[red,  thick, -, dotted] (hf.east)-- (dc.west);
   \draw[red,  thick, -, dotted] (hh.east)-- (dd.west);   
\end{tikzpicture}
\caption{Un-allocation.}
\label{fig:Ex4P5}
\end{figure}
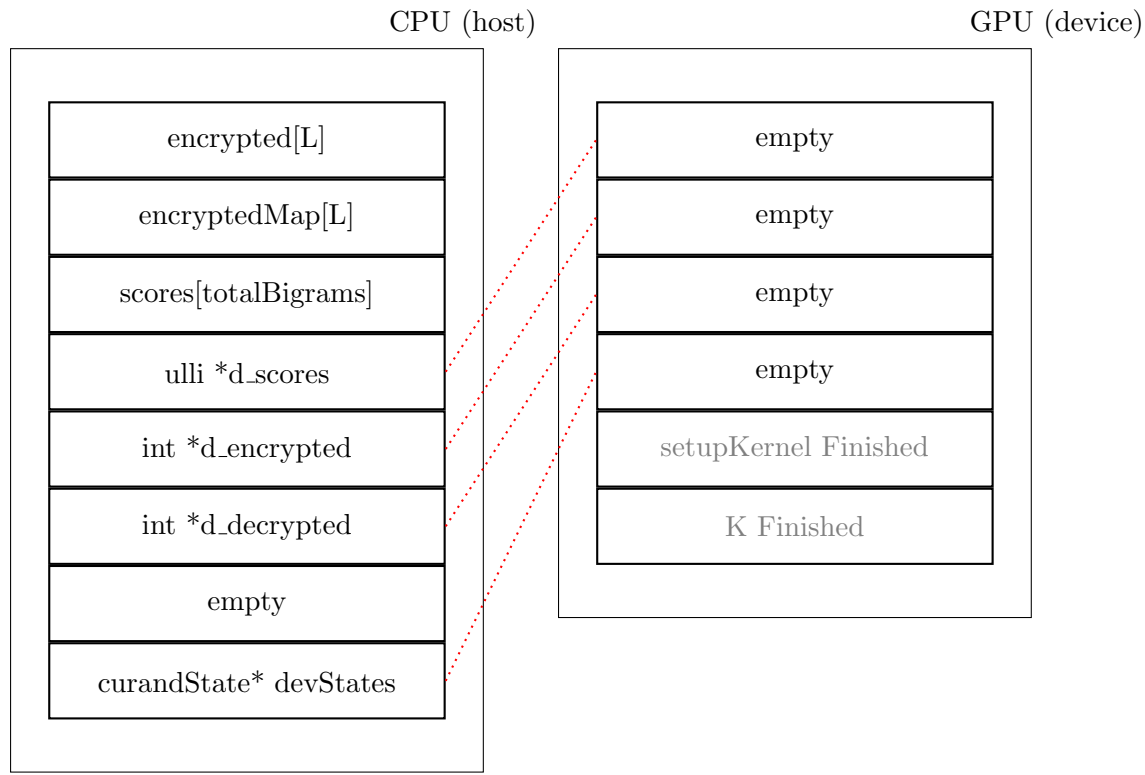

Once the host integer array \textbf{decrypted} has been populated with the local optimums yielded by the device threads, we pick the best one and announce it as a solution. However, before doing that, we should make sure that all the threads had finished their hill climbing routines by synchronizing the threads (see line 143 from Listing \ref{list:multiHill}). We are ready to un-allocate the used memory blocks (see Figure \ref{fig:Ex4P5}).

We consequently un-allocate all the device memory blocks we have used throughout the automatic cryptanalysis routines (see lines 165-168 from Listing \ref{list:multiHill}). However, as shown in line 169, we further need to free the dynamically allocated memory blocks from the host as well. Since the array \textbf{decrypted} was created dynamically (line 121) we further clean it by calling the host \textbf{delete} operator.

Now, let's go back and inspect the new version of the kernel \textbf{K}. We start with copying the original encrypted message to the GPU thread (lines 52-54 from Listing \ref{list:multiHill}). Then, we extract the unique thread identification number (line 55), initialize and load the seed from the device global memory (line 56), initialize some helper variables (lines 58-62) and initiate the hill climbing routines (lines 64-100). 

Each climbing try starts with resetting the \textbf{delta} variable to 0. Then, we generate a random float number by using the thread seed (line 66) and further convert it to an integer number in the interval [0,26)(line 67). We repeat the process to get another distinct integer number in the same interval (lines 68-72). Those numbers are saved in two variables named \textbf{leftLetter} and \textbf{rightLetter}. 

At lines 74-88 we interchange all occurrences of the letter \textbf{leftLetter} inside the encrypted text with the letter \textbf{rightLetter} and vice versa. The variable \textbf{delta} is tracing the score change when the old bigram (line 78) is interchanged with the new one (line 89). Indeed, if the overall sum of deltas is greater than zero, then the newly created candidate is better than the old one and we update it accordingly (that's why we have created a local copy of the encrypted text in the first place). This is visible at lines 92-100. We further save the best thread candidate to the device global memory, so we can later extract it by the host.

Most of the routines inside the \textbf{main} function were already discussed in the previous sections. At the end of the function, at lines 148-158, we further extract the best candidate among all thread final candidates and print it to the terminal (line 162).

\begin{figure}[H]
\centering
\includegraphics[width=300px]{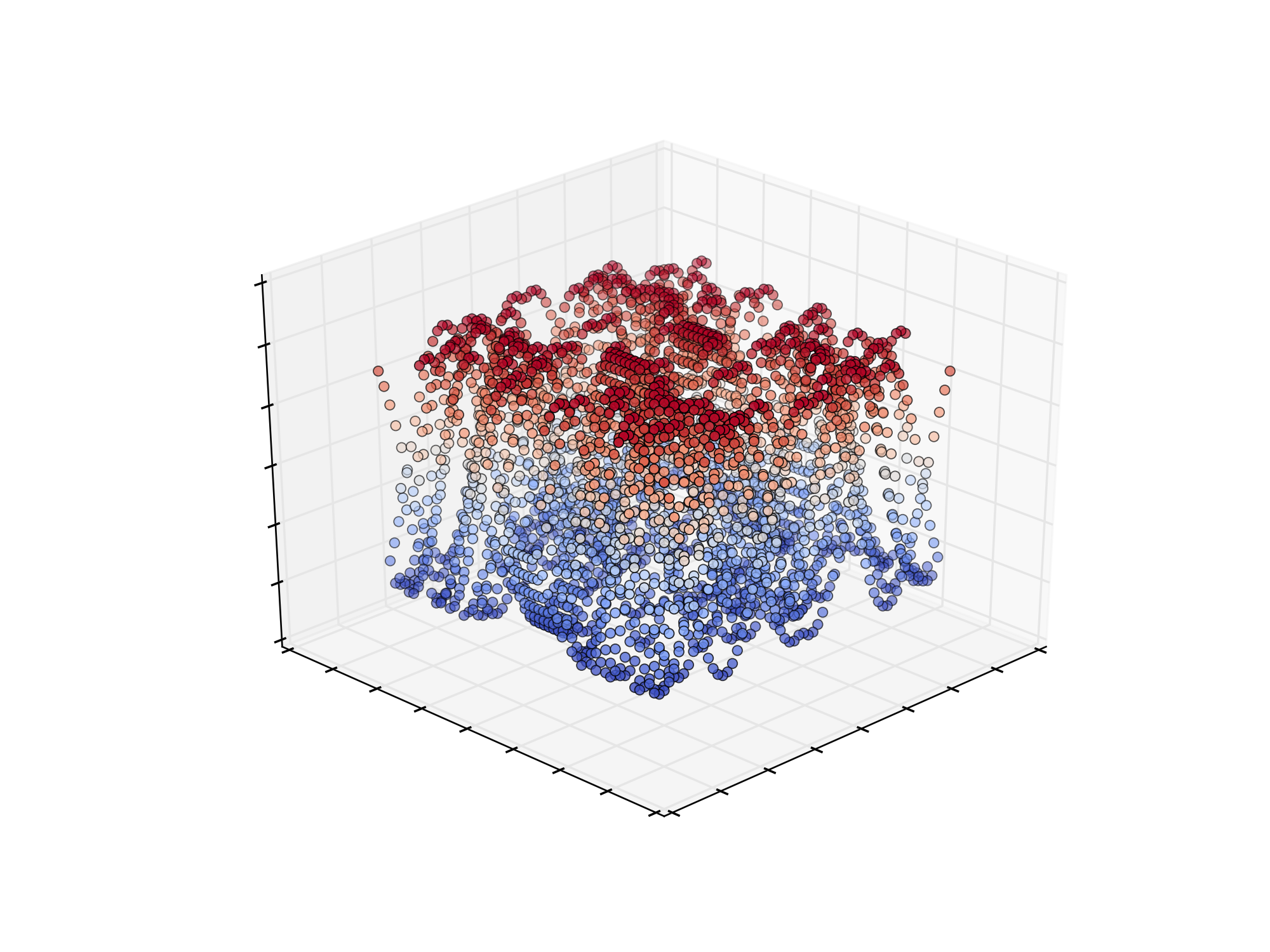} 
\caption{Example of CUDA parallel hill-climbing discretization}
\label{fig:ExampleTopologyDiscrete} 
\end{figure}

The parallel hill climbing approach can be illustrated by the visualization shown in Figure \ref{fig:ExampleTopologyDiscrete}. Each circle represents a local maximum, i.e. a peak, while the color of the circle corresponds to the peak altitude -- lower values are colored in cold colors (nuances of blue), while the higher values are represented with hot colors (nuances of red).

Each thread of the CUDA device is to be attracted by some of those peaks. Then, the peaks are further collected to make measurements of their exact altitude in order to announce the best candidate.

Now, let's compile and run the source code given in Listing \ref{list:multiHill}. Can you decrypt the provded encrypted text? Who is the original author of the plain text? 

Finally, you could try to decrypt the message given in Listing \ref{list:FirstMASCipher} as well. What was the success ratio, i.e. the number of times you execute the program versus the number of times the global solution has been found? 

\section{CUDA memory model and CUDA error handling}
\label{sec:CUDADebug} 

This section briefly discusses a simple way to catch CUDA related errors after a given kernel launch. Besides that, we will take a look into the hierarchy of the CUDA memory model to emphasize its importance.

There are several types of memory blocks inside a given CUDA device. Table \ref{tab:memoryBlocks} gives an overview of these types. A more detailed introduction to the CUDA memory hierarchy could be found in \cite{cudamemory2}.

\begin{table}[h]
	\setlength{\tabcolsep}{12pt}
  \begin{center}
  	\caption{Overview of the types of memory blocks in a CUDA device }
  	\label{tab:memoryBlocks}
    \begin{tabular}{@{} r|c|c|c @{}}      
      \hline
      \textbf{Memory} & \textbf{Encapsulation} & \textbf{Size} & \textbf{Latency} \\
      \hline
      \hline
      Local & Per Thread & \textcolor{red}{Limited} & \textcolor{ForestGreen}{Small}\\   
      Shared & Per Block & \textcolor{Orange}{Mediocre} & \textcolor{Orange}{Medium}\\    
      Global & Per Device & \textcolor{ForestGreen}{Huge} & \textcolor{red}{High}\\ 
	  \hline	
    \end{tabular}
    \end{center}	
\end{table}

The variables defined inside the kernel are normally saved inside the local memory (composed of registers). Let's assume that we have a total of $T$ threads to be launched by some CUDA kernel. Furthermore, each thread requires a total of $X$ bytes to be allocated. So, we have a total of $T*X$ bytes that need to be situated inside the CUDA core registers. If we have a total of $C$ CUDA cores, each having a fixed number of available size of registers of $Y$ bytes, the total available CUDA core register size is $C*Y$. What happen if $T*X>C*Y$? This possible scenario is defined as registers \textbf{spilling} and the extra required memory is allocated on the global memory. However, this greatly reduce the software performance.

During our construction of the CUDA cryptanalysis tools we ignored the memory model. We left that optimization to the CUDA compiler. Now, let's update the CUDA kernel of our previous example to benefit from the \textbf{shared memory} (see Listing \ref{listing:18}).

\begin{lstlisting}[label={listing:18}, caption={Parallel CUDA pseudo-random float number generator kernel, optimized by shared memory utilization}, style=customc, numbers=left,stepnumber=1]
__global__ void K(unsigned long long int *d_scores, int *d_encrypted, int encryptedLen,  curandState* globalState, int *d_decrypted){

	__shared__ unsigned long long int shared_scores[ALPHABET*ALPHABET];

	int* localBufferEncrypted = new int[encryptedLen];
	for (int j=0; j<encryptedLen; ++j)
		localBufferEncrypted[j] = d_encrypted[j];
	int idx = blockIdx.x*blockDim.x + threadIdx.x;
	
	if (threadIdx.x <= 25) {
		for (int j=0; j<ALPHABET; ++j)
			shared_scores[threadIdx.x*ALPHABET + j] = d_scores[threadIdx.x*ALPHABET + j];
	}
	
	__syncthreads();
	
	curandState localState = globalState[idx];

	long long int delta = 0;
	float randF = 0;
	int leftLetter = 0;
	int rightLetter = 0;
	int pair[2];
	
	for (int cycles=0; cycles<CLIMBINGS; ++cycles) {	
		delta = 0;		
		randF = curand_uniform(&localState);
		leftLetter = randF*ALPHABET; 	
		rightLetter = leftLetter;
		while (rightLetter == leftLetter) {
			randF = curand_uniform(&localState);
			rightLetter = randF*ALPHABET; 
		}
		
		for (int j=0; j<encryptedLen-1; ++j)
			{
				pair[0] = localBufferEncrypted[j];
				pair[1] = localBufferEncrypted[j+1];
				delta -= shared_scores[ALPHABET*pair[0] + pair[1]];
				
				if (pair[0] == leftLetter)
					pair[0] = rightLetter;
				else if (pair[0] == rightLetter)
					pair[0] = leftLetter;
					
				if (pair[1] == leftLetter)
					pair[1] = rightLetter;
				else if (pair[1] == rightLetter)
					pair[1] = leftLetter;
					
				delta += shared_scores[ALPHABET*pair[0] + pair[1]];				
			}			
			
			if (delta > 0) {
				for (int j=0; j<encryptedLen; ++j) {
					if (localBufferEncrypted[j] == leftLetter)
						localBufferEncrypted[j] = rightLetter;
					else if (localBufferEncrypted[j] == rightLetter)
						localBufferEncrypted[j] = leftLetter;
				}
			}
		}
	
	for (int j=0; j<encryptedLen; ++j)
		d_decrypted[idx*encryptedLen+j] = localBufferEncrypted[j];
}
\end{lstlisting}

As shown at line 3 we declare an array named \textbf{shared\_scores}, which is going to be visible to all the threads inside the given block. At lines 10-13 we partially redistribute the transfer of the bigram scores from the global memory (d\_scores array) to the shared memory. Having this in mind, we need to synchronize all the threads before stepping into the next source code fragments (line 15). If we fail to do that, there is a great chance for some subset of threads to start reading parts of the shared memory block with undefined values. Then, we proceed with the bigram scoring routines as usual, but this time extracting the scores from the shared memory instead of the global one (lines 39 and 51).

This simple utilization of the device shared memory resulted in considerable speed improvements. For example, during our experiments, an encrypted text with length 471 symbols is decrypted on average in 1.1 seconds by using a mid-range GPU device and global memory only. However, by exploiting the shared memory utilization we did that in less than 0.5 seconds.

In order to squeeze out the best of a given CUDA capable device we need to tune up the thread and number values. Now, introducing the shared memory concept it makes sense to organize the threads into blocks. However, we should pay attention to the technical parameters of the CUDA device. Otherwise, we can end up in the situation with unpredictable semantic errors. Nevertheless, we can catch some of those errors by using the following \textbf{error handle fragment} after each CUDA kernel launch (see Listing \ref{listing:19}):

\begin{lstlisting}[label={listing:19}, caption={Example of error handling in CUDA, optimized by shared memory utilization}, style=customc, numbers=left,stepnumber=1]
cudaError_t error = cudaGetLastError();
	if (error != cudaSuccess) {
		printf("CUDA error: %s\n", cudaGetErrorString(error));
		exit(-1);
	}
\end{lstlisting}

\section{Dynamical vs static GPU memory allocation}
\label{sec:dynamic}

The automatic cryptanalysis tool presented in Listing \ref{list:multiHill} was constructed in a flexible way, which, for example, allows us to re-configure the source code to be applicable for another encrypted text with different length by minimum efforts. However, to achieve this we used a technique which is more common for CPU programming, not GPU, i.e. dynamic memory allocation. 

We can significantly improve the performance of the GPU device if dynamic memory allocation routines are avoided. Listing \ref{list:multiHillStatic} contains the complete CUDA source code of the static implementation of the solver.

This time the need of dynamically memory allocation is avoided by exploiting the C/CUDA preprocessor directives (see lines 7-15). Compile and run the source code in Listing \ref{list:multiHillStatic}. Can you recover the plaintext of the encrypted message? 


\begin{lstlisting}[label={list:multiHillStatic},caption={Final CUDA implementation for automatic cryptanalysis of text encrypted by \ac{mas}, optimized by shared memory utilization}, style=customc, numbers=left,stepnumber=1]
#include <stdio.h>
#include <stdlib.h>
#include <string.h>
#include <cuda.h>
#include <curand_kernel.h>

#define B ((int)16)
#define T ((int)26)
#define THREADS ((int)B*T)
#define CLIMBINGS ((int)5000)
#define ALPHABET ((int)26)
#define totalBigrams ((int)ALPHABET*ALPHABET)

#define encrypted "djrggrygdudrdunkluejqxgahbvhxbixnadjng qniqwdxqaqguaeouludbdndjngqsjngqlnaeoulhdunktqiuqghkbnxtu khxbaqdjntnihkhobgughktsjngqgnordunkxqmruxqgdjqenggqggunk nionkzaqgghzqghktarljduaqhktgdrtbinxdrkhdqobdjqanxqtuiiul rodgrygdudrdunkluejqxghxqxhxqobrgqtinxauoudhxberxengqgnkh llnrkdnidjqduaqhktlhxqxqmruxqtinxqkluejqxukzhkttqluejqxukz"
#define ENCRYPTEDLEN ((int)sizeof(encrypted)-1)

__host__ void extractBigrams(long long int *scores) {
	FILE* bigramsFile = fopen("bigramParsed", "r");
	while(1){
		char tempBigram[2];
		long long int tempBigramScore = 0;
		if (fscanf(bigramsFile, "%s %llu", tempBigram, &tempBigramScore) < 2)
			 { break; } 
		scores[(tempBigram[0]-'a')*ALPHABET + tempBigram[1]-'a'] = tempBigramScore; 
	}
	fclose(bigramsFile);
}

__host__ int getMaxElement(long long int *scores, int length) {
	long long int tempMaxScore = 0;
	int tempIndex = 0;	
	for (int j=0; j<length; ++j) {
		if (scores[j] > tempMaxScore) {
			tempIndex = j;
			tempMaxScore = scores[j];
		} 
	}	
	return tempIndex;
}

__host__ __device__ void demap(int* encrMap) {
	for (int j=0; j<ENCRYPTEDLEN; ++j) {
		printf("%c", encrMap[j] + 'a');
	}
	printf("\n");
}

__host__ long long int candidateScore(int* decrMsg, long long int* scores) {
	long long int total = 0;
	for (int j=0; j<ENCRYPTEDLEN-1; ++j) {
		total += scores[ALPHABET*decrMsg[j] + decrMsg[j+1]];
	}
	return total;
}

__global__ void setupKernel(curandState* state, long seed) {
	int idx = blockIdx.x*blockDim.x + threadIdx.x;
	curand_init(seed, idx, 0, &state[idx]);
}

__global__ void K(long long int *d_scores, int *d_encrypted, curandState* globalState, int *d_decrypted){

	__shared__ long long int shared_scores[totalBigrams];
	
	int localBufferEncrypted[ENCRYPTEDLEN];
	for (int j=0; j<ENCRYPTEDLEN; ++j)
		localBufferEncrypted[j] = d_encrypted[j];
	int idx = blockIdx.x*blockDim.x + threadIdx.x;
	
	if (threadIdx.x <= 25) {
		for (int j=0; j<ALPHABET; ++j)
			shared_scores[threadIdx.x*ALPHABET + j] = d_scores[threadIdx.x*ALPHABET + j];
	}
	
	__syncthreads();
	
	curandState localState = globalState[idx];

	long long int delta = 0;
	float randF = 0;
	int leftLetter = 0;
	int rightLetter = 0;
	int pair[2];
	
	for (int cycles=0; cycles<CLIMBINGS; ++cycles) {	
		delta = 0;		
		randF = curand_uniform(&localState);
		leftLetter = randF*ALPHABET; 	
		rightLetter = leftLetter;
		while (rightLetter == leftLetter) {
			randF = curand_uniform(&localState);
			rightLetter = randF*ALPHABET; 
		}
		
		for (int j=0; j<ENCRYPTEDLEN-1; ++j)
			{
				pair[0] = localBufferEncrypted[j];
				pair[1] = localBufferEncrypted[j+1];
				delta -= shared_scores[ALPHABET*pair[0] + pair[1]];
				
				if (pair[0] == leftLetter)
					pair[0] = rightLetter;
				else if (pair[0] == rightLetter)
					pair[0] = leftLetter;
					
				if (pair[1] == leftLetter)
					pair[1] = rightLetter;
				else if (pair[1] == rightLetter)
					pair[1] = leftLetter;
					
				delta += shared_scores[ALPHABET*pair[0] + pair[1]];				
			}			
			
			if (delta > 0) {
				for (int j=0; j<ENCRYPTEDLEN; ++j) {
					if (localBufferEncrypted[j] == leftLetter)
						localBufferEncrypted[j] = rightLetter;
					else if (localBufferEncrypted[j] == rightLetter)
						localBufferEncrypted[j] = leftLetter;
				}
			}
		}
	
	for (int j=0; j<ENCRYPTEDLEN; ++j)
		d_decrypted[idx*ENCRYPTEDLEN+j] = localBufferEncrypted[j];
		
}

int main() {

	int encryptedMap[ENCRYPTEDLEN];
	
	for (int j=0; j<ENCRYPTEDLEN; ++j)
		encryptedMap[j] = encrypted[j] - 'a';
	
	long long int scores[totalBigrams];	
	extractBigrams(scores);
		
	long long int *d_scores;
	int *d_encrypted, *d_decrypted;
	static int decrypted[ENCRYPTEDLEN*THREADS];
	
	cudaMalloc((void **)&d_scores, sizeof(long long int)*totalBigrams);
	cudaMalloc((void **)&d_encrypted, sizeof(int)*ENCRYPTEDLEN);
	cudaMalloc((void **)&d_decrypted, sizeof(int)*ENCRYPTEDLEN*THREADS);	
	
	cudaError_t error = cudaGetLastError();
	if (error != cudaSuccess) {
		printf("CUDA error (1): %s\n", cudaGetErrorString(error));
		exit(-1);
	}
	
	cudaMemcpy(d_scores, scores, sizeof(long long int)*totalBigrams, cudaMemcpyHostToDevice);
	cudaMemcpy(d_encrypted, encryptedMap, sizeof(int)*ENCRYPTEDLEN, cudaMemcpyHostToDevice);
	
	error = cudaGetLastError();
	if (error != cudaSuccess) {
		printf("CUDA error (2): %s\n", cudaGetErrorString(error));
		exit(-1);
	}
	
	curandState* devStates;
	cudaMalloc(&devStates, THREADS*sizeof(curandState));
	
	error = cudaGetLastError();
	if (error != cudaSuccess) {
		printf("CUDA error (3): %s\n", cudaGetErrorString(error));
		exit(-1);
	}
		
	setupKernel<<<B,T>>>(devStates, time(NULL));	
	cudaDeviceSynchronize();
	
	error = cudaGetLastError();
	if (error != cudaSuccess) {
		printf("CUDA error (4): %s\n", cudaGetErrorString(error));
		exit(-1);
	}
	
	K<<<B,T>>>(d_scores, d_encrypted, devStates, d_decrypted);
	cudaDeviceSynchronize();
	
	error = cudaGetLastError();
	if (error != cudaSuccess) {
		printf("CUDA error (5): %s\n", cudaGetErrorString(error));
		exit(-1);
	}

	cudaMemcpy(decrypted, d_decrypted, sizeof(int)*ENCRYPTEDLEN*THREADS, cudaMemcpyDeviceToHost);
	
	int bestCandidate = 0;
	long long int bestScore = 0;
	
	for (int j=0; j<THREADS; ++j)	{
		long long int currentScore = candidateScore(&decrypted[ENCRYPTEDLEN*j], scores);
		if (currentScore > bestScore) {
			bestScore = currentScore;
			bestCandidate = j;
		}		
	}	
	
	printf("Best candidate score: %llu\n", bestScore);
	demap(&decrypted[ENCRYPTEDLEN*bestCandidate]);		
			
	cudaFree(d_scores);
	cudaFree(d_encrypted);
	cudaFree(d_decrypted);
	cudaFree(devStates);

	return 0;	
}
\end{lstlisting}

\section{Used principles and comparison to other implementations}
\label{sec:principles_comparison}

This Section summarizes the improvements made throughout the tutorial. Then, by following the design principles outlined during the previous sections, a stand-alone CUDA application for automatic cryptanalysis of ciphertexts encrypted by \acl{sct} is presented. At the end the provided GPU routine is  compared with CT2 \cite{esslinger2009cryptool}: a state-of-the-art cryptanalysis tool.

\subsection{Generalization of principles}

Throughout this tutorial, we addressed common issues, which arose from translating a given problem from the domain of classical cryptanalysis to the domain of general-purpose computations on graphics processing units (\textbf{GPGPU}). Step by step, by introducing technical optimization and mathematical insights, we built an efficient  stand-alone framework. Let's summarize the most critical improvements we made:

\begin{itemize}
\item{\textbf{Reducing the overhead caused by the host-device communication (Section \ref{subsec:frequencyAnalysis})}: The first optimization issue was caused by the bandwidth overhead, generated from the communication between the host and the device. This, for example, is illustrated in Listing 7. We pick two different random letters, pass them to the GPU kernel, which logically divides the work to the 325 threads. Then, the best score is fetched and we repeat this process 500 times. However, this creates additional bandwidth overhead as we repeatedly send packets back and forth between the device and the host. To get rid of this undesired behavior, we made the following changes:
	\begin{itemize}
	\item{The \ac{prng} was reallocated  from the CPU to the GPU itself (\textbf{Section \ref{subsec:pseudoRandom}}).}
	\item{The CUDA application was switched to an heuristic version and the host-device bandwidth was optimized. Due to this modification, the hill climbing routine is launched only once (\textbf{Section \ref{sub:standalone}}).}
	\end{itemize}}
	
\item{\textbf{Utilizing a metaheuristic approach}, which is both: highly effective in solving the problem and is efficiently implemented to split the overall work between different GPU cores. Various nuances of metaheuristics were exploited throughout the  tutorial. We first started with \textbf{best neighbor} hill climbing approach (\textbf{Section \ref{subsec:Implement}}). Then, we migrated to the \textbf{better neighbor} metaheuristic (\textbf{Section \ref{sub:standalone}}). This migration was possible by encapsulating the \ac{prng} to the GPU device.}

\item{\textbf{Synchronizing the threads (\textbf{Section \ref{sub:standalone}})}: Synchronization was achieved by using the following techniques:}
	\begin{itemize}
	\item{Each thread starts from a pseudo-random state. Hence, the key space is randomly crawled.}
	\item{Each thread is a stand-alone, i.e. it is capable of performing the full instance of the algorithm entirely by itself.}
	\end{itemize}
	
\item{\textbf{Using faster memory when possible (\textbf{Section \ref{sec:CUDADebug}}):} The most-frequently used memory read operations were migrated to the device shared memory space.}

\item{\textbf{Avoiding dynamic memory allocations (Section \ref{sec:dynamic}).}}
\end{itemize}

\subsection{A stand-alone CUDA application for automatic ciphertext-only attacks against \acl{sct}}
\label{subsec:standAlone}

Following these design principles, the last example (see Listing \ref{list:multiHillStatic}) was slightly modified to be launched on a more complex problem, the cryptanalysis of the Single-columnar transposition \ac{sct} cipher. An introduction, as well as an overview of the state-of-the-art attacks on the \ac{sct} cipher, can be found in the PhD thesis of G. Lasry \cite{lasry2018methodology}. The source code in Listing \ref{list:SingleColumn} is a CUDA implementation for automatic cryptanalysis of encrypted by \ac{sct}.

\begin{lstlisting}[label={list:SingleColumn},caption={Automatic cryptanalysis of ciphertext encrypted by \ac{sct}}, style=customc, numbers=left,stepnumber=1]
#include <stdio.h>
#include <stdlib.h>
#include <string.h>
#include <cuda.h>
#include <curand_kernel.h>

#define B ((int)32)
#define T ((int)32)
#define THREADS ((int)B*T)
#define CLIMBINGS 150000
#define ALPHABET 26
#define totalBigrams ((int)ALPHABET*ALPHABET)
#define CAP ((float)999999.0)

#define ENCRYPTED_T "tteohtedanisroudesereguwocubsoit oabbofeiaiutsdheeisatsarsturesuaastniersrotnesctrctxd iwmhcusyenorndasmhaipnnptmaeecspegdeislwoheoiymreeotb sspiatoanihrelhwctftrhpuunhoianunreetrioettatlsnehtba ecpvgtltcirottonesnobeeeireaymrtohaawnwtesssvassirsrh abapnsynntitsittchitoosbtelmlaouitrehhwfeiaandeitcieg freoridhdcsheucrnoihdeoswobaceeaorgndlstigeearsotoetd uedininttpedststntefoeaheoesuetvmmiorftuuhsurof"
#define ENCRYPTEDLEN ((int)sizeof(ENCRYPTED_T)-1)

#define KEY_LENGTH 30
#define SECTION_CONSTANT ENCRYPTEDLEN/KEY_LENGTH

#define HEUR_THRESHOLD_OP1 50
#define HEUR_THRESHOLD_OP2 70

#define OP1_HOP 4
#define OP2_HOP 2

__host__ void extractBigrams(float *scores) {
	FILE* bigramsFile = fopen("log2", "r");
	while(1){
		char tempBigram[2];
		float tempBigramScore = 0.0;
		if (fscanf(bigramsFile, "%s %f", tempBigram, &tempBigramScore) < 2)
			 { break; } 
		scores[(tempBigram[0]-'a')*ALPHABET + tempBigram[1]-'a'] = tempBigramScore; 
	}
	fclose(bigramsFile);
}

__host__ __device__ void demap(int* encrMap) {
	for (int j=0; j<ENCRYPTEDLEN; ++j) {
		printf("%c", encrMap[j] + 'a');
	}
	printf("\n");
}

__host__ float candidateScore(int* decrMsg, float* scores) {
	float total = 0.0;
	for (int j=0; j<ENCRYPTEDLEN-1; ++j) 
		total += scores[ALPHABET*decrMsg[j] + decrMsg[j+1]];	
	return total;
}

__global__ void setupKernel(curandState* state, unsigned long seed) {
	int idx = blockIdx.x*blockDim.x + threadIdx.x;
	curand_init(seed, idx, 0, &state[idx]);
}

__device__ __host__ void decrypt(int* encrypted, int* key, int* decrypted) {

	int columns[KEY_LENGTH][SECTION_CONSTANT+1];
	int offset = 0;
	int colLength[KEY_LENGTH];
	
	for (int j=0; j<KEY_LENGTH; ++j) {
		colLength[j] = ENCRYPTEDLEN / KEY_LENGTH;
		if (j < ENCRYPTEDLEN % KEY_LENGTH)
			colLength[j]++;
	}
		
	for (int keyPos=0; keyPos < KEY_LENGTH; ++keyPos) {
		offset = 0;
		for (int i=0; i<KEY_LENGTH; ++i)
			if (key[i] < key[keyPos])
				offset += colLength[i];
	
		for (int j=0; j<colLength[keyPos]; ++j)   
			columns[key[keyPos]][j] = encrypted[offset+j];					
	} 
	
	for (int j=0; j<ENCRYPTEDLEN; ++j) 
		decrypted[j] = columns[key[j % KEY_LENGTH]][j / KEY_LENGTH];	
} 

__device__ void swapElements(int *key, int posLeft, int posRight) {	
		if (posLeft != posRight)
		{
			key[posLeft] -= key[posRight];
			key[posRight] += key[posLeft];
			key[posLeft] = key[posRight] - key[posLeft];
		}
}

__device__ void swapBlock(int *key, int posLeft, int posRight, int length) {	
		for (int i=0; i<length; i++) 
			swapElements(key, (posLeft+i)%KEY_LENGTH, (posRight+i)%KEY_LENGTH);
}

__global__ void K(float *d_scores, int *d_encrypted,  curandState* globalState, int *d_decrypted){

	__shared__ float shared_scores[ALPHABET*ALPHABET];
	
	int key[KEY_LENGTH];
	int localDecrypted[ENCRYPTEDLEN];	
	int bestLocalDecrypted[ENCRYPTEDLEN];	
	int idx = blockIdx.x*blockDim.x + threadIdx.x;
	curandState localState = globalState[idx];
	int leftLetter = 0;
	int rightLetter = 0;
	int backupKey[KEY_LENGTH];
	int shiftHelper[KEY_LENGTH];
	int blockStart, blockEnd;
	int l,f,t,t0,n,ff,tt;
	float tempScore = 0.0;
	float bestScore = CAP;
	int j = 0, jj = 0;
	
	if (threadIdx.x == 0) {
		for (j=0; j<ALPHABET;++j)
			for (jj=0; jj<ALPHABET; ++jj)
				shared_scores[j*ALPHABET + jj] = d_scores[j*ALPHABET + jj];
	}
	
	__syncthreads();
	
	for (j=0; j<KEY_LENGTH; ++j) 
		key[j]=j;
		
	for (j=0; j<KEY_LENGTH; ++j) {
		swapElements(key, j, curand_uniform(&localState)*KEY_LENGTH);
	}
		
	for (int cycles=0; cycles<CLIMBINGS; ++cycles) {	
	
		for (j=0; j<KEY_LENGTH;j++)
			backupKey[j] = key[j];
			
		tempScore = 0.0;
		
		int branch = curand_uniform(&localState)*100; 
		
		if (branch < HEUR_THRESHOLD_OP1)
		{
			for (j=0; j<1+curand_uniform(&localState)*OP1_HOP; j++) 
			{
				leftLetter = curand_uniform(&localState)*KEY_LENGTH; 	
				rightLetter = curand_uniform(&localState)*KEY_LENGTH; 
				swapElements(key, leftLetter, rightLetter);
			}						
		}
		
		else if (branch < HEUR_THRESHOLD_OP2)
		{
			for (j=0; j< 1+curand_uniform(&localState)*OP2_HOP;j++)
			{
				blockStart = curand_uniform(&localState)*KEY_LENGTH;
				blockEnd = curand_uniform(&localState)*KEY_LENGTH;
				swapBlock(key, blockStart, blockEnd, 1+curand_uniform(&localState)*(abs((blockStart-blockEnd))-1));
			}
		}
		
		else 
		{
			l = 1 + curand_uniform(&localState)*(KEY_LENGTH-2);
			f = curand_uniform(&localState)*(KEY_LENGTH-1);
			t = (f+1+(curand_uniform(&localState)*(KEY_LENGTH-2)));
			t = t % KEY_LENGTH;
			
			for (j=0; j< KEY_LENGTH;j++)
				shiftHelper[j] = key[j];
				
			t0 = (t-f+KEY_LENGTH) % KEY_LENGTH;
			n = (t0+l) % KEY_LENGTH;
			
			for (j=0; j<n;j++) 
			{
				ff = (f+j) % KEY_LENGTH;
				tt = (((t0+j)%n)+f)%KEY_LENGTH;
				key[tt] = shiftHelper[ff];
			}				
		}			
		
		decrypt(d_encrypted, key, localDecrypted);		
		
		for (j=0; j<ENCRYPTEDLEN-1; ++j) {
			tempScore += shared_scores[ALPHABET*localDecrypted[j] + localDecrypted[j+1]];
		}
	
		if (tempScore < bestScore) {
			bestScore = tempScore;
			for (j=0; j<ENCRYPTEDLEN; ++j) {
				bestLocalDecrypted[j] = localDecrypted[j];
			}
		}		
		
		else 
		{
				for (j=0; j<KEY_LENGTH;j++)
					key[j] = backupKey[j];			
		}
	}
	
	for (j=0; j<ENCRYPTEDLEN; ++j)
		d_decrypted[idx*ENCRYPTEDLEN+j] = bestLocalDecrypted[j];
}

int main() {

	int encryptedMap[ENCRYPTEDLEN];

	for (int j=0; j<ENCRYPTEDLEN; ++j)
		encryptedMap[j] = ENCRYPTED_T[j] - 'a';
	
	float scores[totalBigrams];	
	extractBigrams(scores);
	
	float *d_scores;
	int *d_encrypted, *d_decrypted;
	int* decrypted = new int[ENCRYPTEDLEN*THREADS];
	
	if (cudaSuccess != cudaMalloc((void **)&d_scores, sizeof(float)*totalBigrams))
		printf("Error!\n");
	if (cudaSuccess != cudaMalloc((void **)&d_encrypted, sizeof(int)*ENCRYPTEDLEN))
		printf("Error!\n");
	if (cudaSuccess != cudaMalloc((void **)&d_decrypted, sizeof(int)*ENCRYPTEDLEN*THREADS))
		printf("Error!\n");
	
	cudaMemcpy(d_scores, scores, sizeof(float)*totalBigrams, cudaMemcpyHostToDevice);
	cudaMemcpy(d_encrypted, encryptedMap, sizeof(int)*ENCRYPTEDLEN, cudaMemcpyHostToDevice);
	
	curandState* devStates;
	if (cudaSuccess != cudaMalloc(&devStates, THREADS*sizeof(curandState))) 
		printf("Error!\n");
	
	setupKernel<<<B,T>>>(devStates, time(NULL));	
	cudaDeviceSynchronize();
	
	cudaError_t error = cudaGetLastError();
	if (error != cudaSuccess) {
		printf("CUDA error: %s\n", cudaGetErrorString(error));
		exit(-1);
	}
	
	K<<<B,T>>>(d_scores, d_encrypted, devStates, d_decrypted);
	cudaDeviceSynchronize();
	
	error = cudaGetLastError();
	if (error != cudaSuccess) {
		printf("CUDA error: %s\n", cudaGetErrorString(error));
		exit(-1);
	}

	if (cudaSuccess != cudaMemcpy(decrypted, d_decrypted, sizeof(int)*ENCRYPTEDLEN*THREADS, cudaMemcpyDeviceToHost)) {
		printf("Error!\n");
	}
	
	error = cudaGetLastError();
	if (error != cudaSuccess) {
		printf("CUDA error: %s\n", cudaGetErrorString(error));
		exit(-1);
	}
	
	printf("GPU completed. Calculating the best score ...");
	
	int bestCandidate = 0;
	float bestScore = CAP;
	float* scoreHistory = new float[B*T];
	
	for (int j=0; j<THREADS; ++j)	{
		float currentScore = candidateScore(&decrypted[ENCRYPTEDLEN*j], scores);
		scoreHistory[j] = currentScore;
		if (currentScore < bestScore) {
			bestScore = currentScore;
			bestCandidate = j;
		}		
	}	
	
	printf("Best candidate score: %f\n", bestScore);
	demap(&decrypted[ENCRYPTEDLEN*bestCandidate]);
			
	cudaFree(d_scores);
	cudaFree(d_encrypted);
	cudaFree(d_decrypted);
	cudaFree(devStates);
	delete decrypted;
	delete scoreHistory;
	return 0;
}
\end{lstlisting}

The structure of the GPU implementation is following the same observations made at the beginning of this section (see subsection \ref{subsec:standAlone}). However, since we are now dealing with the different encryption method, some modifications are required. 

The \textbf{decrypt} function of the \ac{sct} cipher is given in lines 58-82. It is visible for both the host and the device. Then, in lines 84-96, we introduce two helper functions \textbf{swapElements()} and \textbf{swapBlock()}. The \textbf{swapElements()} function provides an in-memory flip of two elements, while the \textbf{swapBlock()} function provides an in-memory flip of two continuous blocks, with restriction to overlapping. 

The wrapper of the major CUDA kernel is identical, in terms of logic and structure, to the wrapper of the \ac{mas} analyzer kernel. However, in order to improve the scoring function, some minor changes are introduced, which will significantly improve the success rate of plaintext recovery. As discussed in \cite{lasry2018methodology} and \cite{antal2019cryptanalysis}, the metaheuristic plays an important role in the SCT cryptanalysis. In fact, for larger keys, specifically when combined with a short length of the encrypted message, it is difficult, if not impossible, to predict which metaheuristic strategy is going to be the most effective. However, as discussed in \cite{lasry2018methodology}, there are several search operators that appear to be usually highly efficient:
\begin{itemize}  
\item{An inversion of two elements in the key. Labeled as operation I.}
\item{An inversion of two continuous and non-overlapping blocks in the key. Labeled as operation II.}
\item{A shift of a continuous block inside the key. Labeled as operation III.}
\end{itemize}

From a metaheuristic point of view, populating the algorithm routine with different search operators, each having a non deterministic path, rises many questions. For example, which variables dictate the behavior of a given search operator? How to orchestrate the search operators, i.e. how and when to switch from one search operator to another? In most cases, the answers of all those questions are correlated to the problem we are trying to solve. Hence, there is no single right answer.

Listing \ref{list:SingleColumn} implements a metaheuristic apparatus similar to  the one in \ac{ct2} (version 2.1, stable build 8853.1, 2020) -- an  e-learning software including several applied cryptanalysis components (\cite{kopal2014cryptool, esslinger2009cryptool}):

\begin{itemize}
\item{We have a total of 3 search operators, orchestrated by a variable named \textbf{branch} (line 139). In lines 21-22, two variables (both with the prefix \textbf{HEUR\_THRESHOLD\_OP}) could be tweaked by the user. Let's denote them as $p_1$ and $p_2$, where $p_i$ corresponds to HEUR\_THRESHOLD\_OP\textbf{i}.

The algorithm chooses the first operator with probability $\frac{p_1}{100}$, and the second operator with probability $\frac{p_2 - p_1}{100}$. Thus, the probability of choosing the third operator is $1 - \frac{p_2 - p_1}{100} - \frac{p_1}{100} = 1 - \frac{p_2}{100}$.}

\item{\textbf{Operator I}:  This search operator (see lines 141-149) modifies the key by interchanging at least two elements inside it. Exactly which elements, and how many interchanges should occur, is dictated both by the \ac{prng} and the \textbf{OP1\_HOP} variable restriction -- the total count of interchanges should not exceed this value (see line 4).}

\item{\textbf{Operator II}: This (see lines 151-159) corresponds to operation (2) defined in \cite{lasry2018methodology} -- we just interchange at least two continues and non-overlapping blocks sharing the same length. However, the total count of interchanges should not exceed the value of \textbf{OP2\_HOP}. The user can tweak this value (line 25).}

\item{\textbf{Operator III}: This corresponds to operation (3) defined in the aforementioned work -- we just slide a continues and non-self-overlapping block to the left or right (lines 161-180).} 

\item{\textbf{N-gram log2}: The migration to N-gram log2 is introduced. More details could be found in \cite{nather2005n}. In short, if we have a sentence \enquote{ABC}, it is scored as P(\enquote{AB}) $\cdot$ P(\enquote{BC}), i.e. the overall probability of occurrence of this specific composition of letters, given a pre-calculated word corpora. However, multiplication is a tedious operation. 

Hence, we can use a logarithm, with some arbitrary base, to utilize an additive operation instead. Indeed, let's denote \enquote{AB} as x and \enquote{BC} as y. Thus, for some $N$, the following equations hold true:

\centering{$P(x)P(y)=N$}\\
\centering{$\log_2{\left( P(x)P(y) \right)}= \log_2N$}\\
\centering{$\log_2{P(x)} + \log_2{P(y)}= \log_2N$}

This allows to interchange the multiplication operator with the accumulation operator, when a probability space corpora is used. During our experiments, we used the Google N-gram corpus \cite{googleNgram}.
}

\end{itemize}

We should note that the bigger the key is, the larger values of threads and climbings are needed. Thus, the time required to recover the message increases. 

\subsection{Comparison with state-of-the-art \ac{sct} cryptanalysis}

During our experiments, and by using the above GPU implementation, we were able to successfully recover the plaintexts, corresponding to ciphertexts encrypted by columnar transposition cipher with unknown keys (with lengths no bigger than 40) in less than 20 seconds. For example, given $1120$ CUDA threads (utilizing 1152 CUDA cores), a climbing constant of 15,000 and a ciphertext with 596 symbols, encrypted by a key with length 25, the unknown plaintext was successfully recovered in approximately 5.9 seconds. 

The search space of this problem, given a key of size 25, is $25!$, which is approximately equal to $2^{83}$. Nevertheless, a general-purpose computer equipped with mid-range CUDA capable video card could recover, due to the heuristic nature of the algorithm, the plaintext in less than a minute. Table \ref{tab:CrypToolComparison} compares the CUDA decryption routine with the hill climbing routine in the 2020 version of  \ac{ct2}. We used general-purpose hardware: a video card of NVIDIA 1060, 3 GB, and a CPU of Intel Celeron G1820 @ 2.7 GHz with 2 cores.

\newcommand{\specialcell}[2][c]{\begin{tabular}[#1]{@{}l@{}}#2\end{tabular}}
\newcommand{\specialcellbold}[2][c]{%
  \bfseries
  \begin{tabular}[#1]{@{}l@{}}#2\end{tabular}%
}
\begin{table}[H]
  \begin{center}
  \caption{Elapsed time comparison between CUDA and the hill-climbing solver in  CT 2.1 (build 8853.1)}
  \label{tab:CrypToolComparison}
    \begin{tabular}{lllllll}
    \toprule
    \specialcellbold{Key\\size} &
    \specialcellbold{Search\\space} &
    \textbf{Threads} &
    \textbf{Iterations} &
    \specialcellbold{CUDA\\time} &
    \specialcellbold{CrypTool\\time} &
    \specialcellbold{Speed-up\\factor}\\
    \midrule
      $10$ & $2^{21}$ & $1120$ & 15,000 & 5.5s & 27m & 294\\
      $15$ & $2^{40}$ & $1120$ & 15,000 & 6.5s & 27m & 249\\
      $20$ & $2^{61}$ & $1120$ & 15,000 & 5.5s & 28m & 305\\
      $25$ & $2^{83}$ & $1120$ & 15,000 & 5.9s & 28m & 285\\
      $30$ & $2^{107}$ & $1120$ & 30,000 & 10.7s & 55m & 308\\
      $35$ & $2^{132}$ & $1120$ & 100,000 & 46.2s & 3h 11m & 248\\
    \bottomrule
    \end{tabular}
  \end{center}	
\end{table}


The times shown in Table \ref{tab:CrypToolComparison} correspond to the time needed for the given job to be completed (the time needed for all the threads to complete all the iterations). The numbers do not correspond to the first valid decryption of the encrypted text. The success ratio of the key recovery routine is  entangled to the choice of metaheuristic, search operators or other parameters, which could orchestrate the trace direction through the search space. It would not be fair to  compare the times needed for a given job to be completed, without consideration of the success ratio. Having this in mind, by examining the source code of the CrypTool hill climbing routine (CHC), we have further tuned the CUDA routine to exactly match the search operators and the magic constants to be found throughout CHC. This perfectly synchronized the success ratio of CUDA routine and CHC. However, it  did not affect the time needed for completion of the CUDA algorithm. In overall, the CUDA routine is approximately 250 times faster than the CHC. During the comparison stage, we used the following options: Action: Decrypt, Read in: by row, Permutation: by column, Read out: by column, Function type: N-grams log 2, N-gram size: 2, Language: English.

For more information regarding cryptanalysis of columnar transposition ciphers, we recommend section 5 of Lasry's thesis (\cite{lasry2018methodology}), the paper devoted to breaking historical ciphers from the Biafran war (\cite{bean2020eavesdropping}), as well as the research of cryptanalysis of columnar transposition ciphers with long keys provided in \cite{lasry2016cryptanalysis}.

After completing this tutorial, you might want to look at the attacks presented in \cite{combes1,combes2,combes3,combes4, combes5} against the Sigaba cipher machine (\cite{savard1999ecm}) which was used by the United States for message encryption during the World War II. In the series of posts, Stuart Combes  explains the details of his attack based on the improved method of Stamp and Chan (2007) (\cite{stamp2007sigaba}). By exploiting the fact that most of the computations needed for breaking Sigaba can be parallelized, the attack was practially implemented using CUDA.  Understanding of this attack and how it was implemented can further deepen your knowledge and is a nice example of how CUDA could be used for cryptanalysis of a more enhanced cipher.

\section{Summary}
\label{sec:summary}

This tutorial outlined how multi-core GPU devices can also be beneficial to solve problems related to classical cryptanalysis. Starting from a complete blank project, we built up and efficiently implemented a full CUDA-based tool, which is able to automatically decrypt ciphertexts encrypted by \ac{mas} or columnar transposition ciphers. 

Some major questions raised in the process of designing a CUDA-based cryptanalysis tool  are:

\begin{itemize}
\item{Are we going to use \ac{prng}s? If so, should we initialize them on the device?}

\item{What metaheuristic are we going to apply? Are we going to use a hill climbing approach (straightforwardly accepting the first better candidate)? Or maybe regular neighborhood search defined by some search operator $I$, by always choosing the best neighbor? Or some alternative metaheuristic such as simulated annealing \cite{van1987simulated} and tabu search \cite{glover1998tabu}? }

\item{What is the score, which we are going to use? Is it going to be single score (just bigrams) or multi objective score (maybe, bigrams plus trigrams plus pentagrams)? What is the size of the data related to the cost function, that we need to upload on the device? Is it small enough to be kept inside the registers? If not -- is it small enough to be kept inside the shared memory? If not, maybe we can reshape the blocks or  decrease the threads inside a given block, to further increase our chances to utilize the shared memory?}

\item{How often do we need to transfer data back and forth between the host and the device? Maybe, we can optimize the decryption process to keep such transfers at minimum?}

\item{How are we going to logically organize the threads? Are they going to be entirely undependable or mutually working on the same distinct subset of a given problem? How often do they need to be synchronized?}
\end{itemize} 

Once we have answered the major questions, we can make a sketch of our implementation. During this sketching we can catch a bunch of design errors, which are not so easy to be revealed at the time of the initial design.  Furthermore, by using diagrams we can proactively design the role and scope of each array and function. Moreover, we can estimate the overhead of our draft architecture and try to optimize our method before the actual implementation.  

\paragraph{Acknowledgement}
We want to thank George Lasry for his valuable feedback and suggestions. We would also like to thank Nils Kopal and Armin Krauss for their comments regarding the columnar transposition hill climbing routine implemented in the 2020 version of \ac{ct2}. Furthermore, we are grateful to Vasily Mikhalev for his detailed revision and practical suggestions, which lead to a noticeable improvement of this tutorial.

\newpage
\bibliography{refs.bib}
\bibliographystyle{apalike}


\newpage


\tableofcontents

\newpage
\listoffigures

\newpage
\listoftables

\newpage
\lstlistoflistings

\newpage
\printacronyms[]

\end{document}